\documentclass[prd,aps,floatfix,preprint,nofootinbib]{revtex4}
\usepackage{amsmath,amssymb,amsbsy,color}
\usepackage{longtable}
\usepackage{lscape}
\usepackage{epsfig}
\usepackage{graphicx}
\usepackage{amssymb}
\usepackage{epstopdf}
\newcommand{\msbar}{\overline {\rm MS}}
\newcommand{\SF}{Schr\"odinger functional\ }
\newcommand{\no}{\nonumber}

\begin{document}
 \vspace*{-20mm}
 \begin{flushright}
  \normalsize
  HUPD-1308,\
  KANAZAWA-13-08,\
  KEK-CP-291
 \end{flushright}
\title{
 Running coupling constant and mass anomalous dimension of six-flavor
 SU(2) gauge theory
}

\author{M.~Hayakawa$^{a}$}
\author{K.-I.~Ishikawa$^{b}$}
\author{S.~Takeda$^{c}$}
\author{N.~Yamada$^{d,e}$}
\email[]{norikazu.yamada@kek.jp}

\affiliation{
$^a$ Department of Physics, Nagoya University, Nagoya 464-8602, Japan\\
$^b$ Department of Physics, Hiroshima University, Higashi-Hiroshima
     739-8526, Japan\\
$^c$ School of Mathematics and Physics, College of Science and
     Engineering, Kanazawa University, Kakuma-machi, Kanazawa, Ishikawa 
     920-1192, Japan\\
$^d$ KEK Theory Center, Institute of Particle and Nuclear Studies, High
     Energy Accelerator Research Organization (KEK), Tsukuba 305-0801,
     Japan\\ 
$^e$ School of High Energy Accelerator Science, The Graduate University
     for Advanced Studies (Sokendai), Tsukuba 305-0801, Japan
}

\date{\today}

\begin{abstract}
 In the exploration of viable models of dynamical electroweak symmetry
 breaking, it is essential to locate the lower end of the conformal
 window and know the mass anomalous dimensions there for a variety of
 gauge theories.
 We calculate, with the Schr\"odinger functional scheme, the running
 coupling constant and the mass anomalous dimension of SU(2) gauge
 theory with six massless Dirac fermions in the fundamental
 representation.
 The calculations are performed on $6^4$ - $24^4$ lattices over a
 wide range of lattice bare couplings to take the continuum limit.
 The discretization errors for both quantities are removed
 perturbatively.
% , and this removal turns out to increase a possible value of infrared
% fixed point and the mass anomalous dimension.
 We find that the running slows down and comes to a stop at
 $0.06 \lesssim 1/g^2 \lesssim 0.15$  where the mass anomalous
 dimension is estimated to be $0.26 \lesssim \gamma^*_m \lesssim 0.74$.
\end{abstract}

\keywords{Lattice Gauge Theory; LHC.}
\pacs{12.38.Gc\vspace{-0ex}}
\maketitle

\section{Introduction}
\label{sec:introduction}

A model of dynamical electroweak symmetry breaking, called technicolor
(TC), offers a natural explanation to hierarchies present in the
standard model
(SM)~\cite{Weinberg:1975gm,Eichten:1979ah,Chivukula:2000mb}.
This class of model could potentially suffer from several serious
problems including those associated with the
$S$ parameter~\cite{Peskin:1990zt}, flavor changing neutral
current~\cite{Chivukula:2010tn}, and relatively light Higgs
mass~\cite{Aad:2012tfa}.
All those problems are, however, expected to disappear if an underlying
gauge theory follows ``walking dynamics''~\cite{Holdom:1981rm}.
By walking dynamics we mean that the renormalized gauge coupling runs at
strong coupling only slowly over a wide range of the energy scale.
Thus, a theory behaves as nearly scale invariant and as a strong coupled
gauge theory.
TC models possessing such a property are called walking technicolor
(WTC).
In order for a WTC scenario to work and avoid the problems above,
another important feature is required, a large mass anomalous dimension
$\gamma_m\sim $O(1) at the (nearly) conformal region; otherwise it
fails, at least, to reproduce the observed masses of the standard model
fermions.
$\gamma_m$ is the $O(g^2)$ quantity in perturbation theory; thus
$\gamma_m\sim$ O(1) can be confirmed only by nonperturbative methods.

WTC consists of fermions having a vectorlike coupling to gauge fields
and hence is tractable on the lattice.
%Since the seminal work by Appelquist, Fleming and
%Neil~\cite{Appelquist:2007hu,Appelquist:2009ty} was published, 
So far, a lot of effort has been made on the lattice to locate the
lower end of the conformal window for various gauge systems, to gain a
quantitative feature of walking dynamics and, importantly, to find the
best candidate for
WTC~\cite{Appelquist:2007hu,Appelquist:2009ty,Appelquist:2010xv,Fodor:2009wk,Deuzeman:2008sc,Miura:2011mc,Miura:2012zqa,Shamir:2008pb,DeGrand:2009hu,Jin:2009mc,DeGrand:2009mt,Hasenfratz:2009ea,Hasenfratz:2010fi,Cheng:2013eu,Ohki:2010sr,Lin:2012iw,Itou:2012qn,Hietanen:2008mr,Hietanen:2009az,DelDebbio:2009fd,Bursa:2009we,Nagai:2009ip,Kogut:2010cz,Aoki:2012ve,Ishikawa:2013wf,Ejiri:2012rr,Bursa:2010xn,Karavirta:2011zg,Voronov:2012qx,Hayakawa:2010yn,Hayakawa:2012gf,Hayakawa:2013maa}.\footnote{For
earlier works on many flavor QCD, see, for example,
Refs.~\cite{Fukugita:1987mb,Brown:1992fz,Iwasaki:2003de}.}
In our previous work, the scale dependence of the gauge coupling
constant of ten-flavor QCD ($N_c$=3 and $N_f$=10) is investigated on the
lattice, which provides an evidence of the infrared fixed point
(IRFP)~\cite{Hayakawa:2010yn}.
In this work, we focus on the two-color QCD with six massless
Dirac fermions ($N_c=2$ and $N_f$=6) in the fundamental representation.

The lattice determination of the running coupling of this theory was
performed by three
groups~\cite{Bursa:2010xn,Karavirta:2011zg,Voronov:2012qx}, all of which
adopt the Schr\"odinger functional (SF)
scheme~\cite{Luscher:1992an,Sint:1998iq}.
In Ref.~\cite{Bursa:2010xn} the calculations are carried out with the 
unimproved Wilson fermion.
The continuum limit is attempted with a constant and linear fits
using data obtained on $(L/a)^4=8^4$ - $16^4$ lattices.
With the constant fit, the renormalized coupling appears to reach a fixed
point when $g^2 > 4.02$, although the data do not exclude the
possibility of the absence of the fixed point in the range of couplings
they studied because of large uncertainties.
They also reported that the mass anomalous dimension at the possible
fixed point is larger than 0.135 and that it can be as large as unity at
the largest coupling ($g^2=5.52$) in their calculation.

In Ref.~\cite{Karavirta:2011zg}, the $O(a)$-improved Wilson-clover
action is used on $(L/a)^4=6^4$ - $16^4$ lattices for the running
coupling and on $(L/a)^4=6^4$ - $20^4$ for the running mass.
After the continuum extrapolation linear in $a^2$, it is found that
the running of the renormalized coupling begins to slow down at
$g^2\sim 5$ compared to the two-loop perturbation theory and the $\beta$
function eventually becomes consistent with zero at $g^2\sim$12, which
is obviously larger than that of Ref.~\cite{Bursa:2010xn}.
Under the assumption that the coupling reaches the IRFP, the mass
anomalous dimension at the IRFP is estimated to be $\gamma^*\gtrsim
0.25$.

In Ref.~\cite{Voronov:2012qx}, the status of their ongoing calculation
of the running coupling using the stout-smeared Wilson fermion action is
reported.
One remarkable feature is that they succeed to explore the large
renormalized coupling up to $g^2\sim 35$.
The continuum limit has not yet been taken, but from the behavior of the
coupling data the authors infer the absence of IRFP, which is in
contrast to the previous two works.

In this work, we use the \SF scheme to calculate the running coupling
and mass.
The main differences from the previous works are the perturbative
improvement of the discretization error and the inclusion of the larger
lattice $24^4$.
After our preliminary result was reported in
Ref.~\cite{Hayakawa:2012gf}, the statistics were substantially increased
especially at our largest lattice ($24^4$).
Based on the numerical results, we argue that the data is consistent
with the presence of IRFP.
The same argument is derived in our recent study on
spectroscopy~\cite{Hayakawa:2013maa}.

The paper is organized as follows.
In Sec.~\ref{sec:ptanalysis}, we recall the results in perturbation
theory.
In Sec.~\ref{sec:remarks}, the reason why walking theory with large
anomalous dimension is necessary is reviewed.
The \SF scheme is briefly explained in Sec.~\ref{sec:parameter}.
In Secs.~\ref{sec:analysis-g2} and \ref{sec:analysis-mass}, we describe
the analysis method and present the numerical results of the running
coupling and mass, respectively.
In Sec.~\ref{sec:discussion}, we summarize our work and discuss the
possible loophole in our argument and the future perspective.
Some details and tables of numerical results are collected in the
Appendices.

\section{perturbation theory}
\label{sec:ptanalysis}

In this section, we examine the perturbative predictions of the lower
end of the conformal window, the value of the fixed point, and the mass
anomalous dimension at the fixed point by adopting the $\msbar$
scheme~\cite{Ryttov:2010iz,Pica:2010xq,Ryttov:2013hka}.

\subsection{Fixed point}

We define the $\beta$ function by
\begin{eqnarray}
     \beta(g^2(L))
 &=& L\,\frac{\partial\,g^2(L)}{\partial L}
  =  b_1\,g^4(L)+b_2\,g^6(L)+b_3\,g^8(L)+b_4\,g^{10}(L)+\cdots,
\label{eq:betafunc}
\end{eqnarray}
where $g(L)$ is the gauge coupling renormalized at length scale $L$.
For SU($N_c$) gauge theory with $N_f$ flavors in the fundamental
representation, the first four coefficients are known in the MS (and
$\msbar$) scheme~\cite{vanRitbergen:1997va} (see
Appendix~\ref{sec:four-loop-coefficients}).
Provided that two coupling constants defined in two arbitrary schemes,
$g_1$ and $g_2$, are related as
\begin{eqnarray}
 g_1^2 = g_2^2\ [ 1 + c_1 g_2^2 + O(g_2^4)\, ],
\end{eqnarray}
the first two coefficients $b_1$ and $b_2$ are proven to be
scheme independent.
We can also see that, if $g_1$ is the single-valued function of $g_2$,
the existence of the IRFP is scheme independent.
For $N_c=2$ and $N_f=11$, $b_1$ vanishes and $b_2$ is negative.
Thus, asymptotic freedom is lost for $N_f \ge 11$.
The perturbative predictions for the IRFP, $g^2_{\rm FP}$, in SU(2)
gauge theory with $N_f$ flavors are summarized in
Table~\ref{tab:p-IRFP}.
%%%%%%%%%%%%%%%%%%%%%%%
\begin{table}[tb]
\centering
\begin{tabular}{c|cccccc}
 $N_f$ & 5 & 6 & 7 & 8 & 9 & 10 \\
\hline
 Two-loop & - & 143.56 & 35.59 & 15.79 & 7.48 & 2.90 \\
 \hline
 Three-loop $\msbar$ & 38.10 & 20.68 & 13.25 & 8.65 & 5.26 & 2.47 \\
 \hline
 Four-loop $\msbar$ & - & 30.10 & 15.21 &  9.55 & 5.58  & 2.52\\
\end{tabular}
\caption{The perturbative predictions for the IRFP, $g_{\rm FP}^2$, in
 the $\msbar$ scheme for SU(2) gauge theory.
%, and nontrivial UV fixed point in $\msbar$ scheme.
 }
\label{tab:p-IRFP}
\end{table}
%%%%%%%%%%%%%%%%%%%%%%%

It should be noted that within the perturbation theory the existence of
IRFP is mainly determined by the sign of the highest order term
considered in the $\beta$ function.
For example, at the two-loop (three-loop) approximation, $b_2 < 0$ for
$N_f \ge 6$ ($N_f \ge 4$), and the IRFP exists in the same $N_f$ region.
This is also true for SU(3) gauge theory as shown in
Table~\ref{tab:p-IRFP-su3}.
%%%%%%%%%%%%%%%%%%%%%%%
\begin{table}[tb]
\centering
\begin{tabular}{c|cccccccc}
 $N_f$ & 7 & 8 & 9 & 10 & 12 & 14 & 16\\
\hline
 Two-loop & - & - & 65.80 & 27.74 & 9.47 & 3.49 & 0.52\\
 \hline
 Three-loop $\msbar$ & 30.88 & 18.40 & 12.92 &  9.60 & 5.46 & 2.70 & 0.50\\
 \hline
 Four-loop $\msbar$ & - & 19.47 & 13.47 & 10.24 & 5.91 & 2.81 & 0.50\\
\end{tabular}
\caption{The perturbative predictions for the IRFP, $g_{\rm FP}^2$, in
 the $\msbar$ scheme for SU(3) gauge theory .}
\label{tab:p-IRFP-su3}
\end{table}
%%%%%%%%%%%%%%%%%%%%%%%

Since $b_3$ and higher order terms depend on the scheme, it is possible
that the sign of $b_3$ depends on the scheme, which means that
the smallest number of flavors having the IRFP, {\it i.e.}, the lower
end of the conformal window, depends on the scheme.
Thus, at least, the analysis through the third order cannot give
reliable information about the existence of the IRFP.
With the fourth order term, one could discuss the convergence of the
perturbative series.
From the tables, it is seen, in general, that the difference of
$g^2_{\rm FP}$ from the three-loop and four-loop analyses is reasonably
small ($\lesssim$15 \%) except for $N_f=6$ in SU(2) gauge theory, where
the IRFP increases by 50 \%.

However, even if $g^2_{\rm FP}$ looks reasonably convergent, the IRFP is
not observed in lattice studies of SU(3) gauge theory with eight
flavors.
This fact is interpreted in two ways: either the IRFP does not exist or
the IRFP exists, but is too large to observe.
Since the perturbative prediction of the IRFP in SU(2) gauge theory with
six flavors is large and does not show plausible convergence, it would
be difficult to draw a definite conclusion on the existence of the IRFP.
Indeed, that is what is encountered in the previous
works~\cite{Bursa:2010xn,Karavirta:2011zg,Voronov:2012qx}.

\subsection{Mass anomalous dimension}

In the following discussion, we implicitly assume that the
renormalization scheme under consideration respects chiral symmetry.
We define the  mass, flavor-singlet scalar density, and flavor
non-singlet pseudoscalar density operators renormalized at length scale
$L$ by $\overline{m}(L)$, $S_R(L)$, and $P^a_R(L)$, respectively, as
follows:
\begin{eqnarray}
 \overline{m}(L) &=& Z_m(L)\, m_0,\\
 S_R(L)          &=& Z_S(L)\, S_0,
 \label{eq:psibarpsi}
 \\
 P^a_R(L)          &=& Z_{P}(L)\, P^a_0,
\end{eqnarray}
where the quantities with the subscript ``0'' denote the bare
quantities.
Then, the partially conserved axial-vector current (PCAC) relations lead
to
\begin{eqnarray}
 Z_m(L)=\frac{1}{Z_{P}(L)}=\frac{1}{Z_S(L)}.
\end{eqnarray}
In the following, we extract the mass anomalous dimension $\gamma_m$
from the scale dependence of $Z_{P}(L)$.

We define the mass anomalous dimension by
\begin{eqnarray}
     \gamma_m(g^2(L))
 &=& \frac{L}{\overline{m}(L)}
     \frac{\partial\,\overline{m}(L)}{\partial L}
  =  \frac{L}{Z_m(L)}
     \frac{\partial Z_m(L)}{\partial L}
  = -\frac{L}{Z_{P}(L)}\frac{\partial Z_{P}(L)}{\partial L}
\no\\  
  &=&  d_1\,g^2(L)\,+d_2\,g^4(L)+d_3\,g^6(L)+d_4\,g^8(L)+\cdots.
\label{eq:dif-eq-zp}
\end{eqnarray}
The first four coefficients in the MS (and the $\msbar$) scheme
are available~\cite{Vermaseren:1997fq}, and the explicit expressions are
found in Appendix~\ref{sec:four-loop-coefficients}.
In this case, only $d_1$ is scheme independent.
If an IRFP exists, the mass anomalous dimension at the fixed point
is also scheme independent.

Now let us see the perturbative prediction.
$\gamma_m^*$ turns out to drastically change by including higher order
terms in $\beta$ and $\gamma_m$.
For example, with the three-loop value of the fixed point
($g^2_{\rm FP}=20.68$), the mass anomalous dimensions including one-,
two-, three- and, four-loop corrections take 0.59, 1.19, 0.93, and
$-$0.26, respectively.
With the four-loop value of the fixed point ($g^2_{\rm FP}=30.10$), they
are 0.86, 2.14, 1.31, and $-$4.02.
Thus, no stable prediction is obtained from perturbative analysis.

\section{Phenomenological requirement}
\label{sec:remarks}

In this section, we review that walking dynamics with a large anomalous
dimension is needed in viable TC models.
One crucial point in the WTC (or extended TC) scenario is how large
quarks' and leptons' masses can be achieved~\cite{Chivukula:2010tn},
where the scalar condensate of the technifermions plays a key role.
The fermion masses in the SM should be given by the RG invariant
condensate
\begin{eqnarray}
     m_{\rm SM,f}
 &=& \frac{C_S^{\rm X}(1/\mu)}{M_{\rm ETC}^2}\
     \langle\ S_R^{\rm X}(1/\mu)\ \rangle,
 \label{eq:sm-fermion-mass}
\end{eqnarray}
where $\mu$ is the renormalization scale, the superscript ``X''
labels the renormalization scheme chosen and $\langle \cdots \rangle$
denotes the vacuum expectation value.
The coefficient $C_S^{\rm X}(\mu)$ is the dimensionless coefficient and
is of $O(1)$ at $\mu=M_{\rm ETC}$.
Its precise value depends on an explicit ETC model.
Equation~(\ref{eq:sm-fermion-mass}) does not depend on the scheme nor
the scale.\footnote{In this argument, QCD and any other interactions and
the corresponding scale dependences are ignored.}

From Eq.~(\ref{eq:psibarpsi}), the condensate at $\mu=M_{\rm ETC}$ can
be written in terms of the condensate at any other scale as
\begin{eqnarray}
     \langle\ S_R^{\rm X}(1/M_{\rm ETC})\ \rangle
 &=& \frac{Z_S^{\rm X}(1/M_{\rm ETC})}{Z_S^{\rm X}(a)}\,
     \langle\ S_R^{\rm X}(a)\ \rangle\ ,
\end{eqnarray}
where the lattice cutoff $a$ is chosen as an example.
The first factor on the rhs. describes the running of the scalar
bilinear operator, or equivalently, the running of the renormalized
mass, and is calculated in the following sections with the SF scheme.
Thus X is set to SF.
The second factor $\langle\ S_R^{\rm X}(a)\ \rangle$ can be determined
on the lattice in the lattice regularization scheme and thus needs a
finite renormalization
\begin{eqnarray}
     S_R^{\rm SF}(a)
 &=& C_S^{\rm SF-Lat}(a)\ S_R^{\rm Lat}(a),
\end{eqnarray}
connecting the SF and the lattice schemes,
The factor $C_S^{\rm SF-Lat}(a)$ can be calculated nonperturbatively as
well on the lattice, although we do not calculate it in this paper.

After all, the masses of fermions in the SM is expressed as
\begin{eqnarray}
     m_{\rm SM,f}
 &=& \frac{C_S^{\rm SF}(1/M_{\rm ETC})}{M_{\rm ETC}^2}\
     \frac{Z_S^{\rm SF}(1/M_{\rm ETC})}{Z_S^{\rm SF}(a)}\
     C_S^{\rm SF-Lat}(a)\ \langle\ S_R^{\rm Lat}(a)\ \rangle\\
 &=& \frac{C_S^{\rm SF}(1/M_{\rm ETC})}{M_{\rm ETC}^2}\
     \frac{Z_S^{\rm SF}(1/M_{\rm ETC})}{Z_S^{\rm SF}(a)}\
     C_S^{\rm SF-Lat}(a)\
     \frac{\langle\ S_R^{\rm Lat}(a)\ \rangle}{f_{\pi_{\rm T}}^3}
     \times (246 {\rm GeV})^3\, ,
 \label{eq:smf-mass}
\end{eqnarray}
where in the last equation the condensate is normalized by the
technipion decay constant $f_{\pi_{\rm T}}=$246 GeV.
The mass anomalous dimension is required in estimating the second factor
$Z_S^{\rm SF}(1/M_{\rm ETC})/Z_S^{\rm SF}(a)$ and is the main subject of
this work.

In the classical TC model, the scalar condensate at $\mu=M_{\rm TC}$ is
estimated to be $M_{\rm TC}^3$, and the other factors are naturally
assumed to be of $O(1)$.
Assuming $M_{\rm ETC}\sim$ 1,000 TeV and $M_{\rm TC}\sim$ 1 TeV,
$m_{\rm SM,f}$ ends up with $M_{\rm TC}^3/M_{\rm ETC}\sim$ 1 MeV,
and hence even the strange quark mass cannot be explained.

In the WTC model, it is expected that the huge enhancement of the second
factor in Eq.~(\ref{eq:smf-mass}) occurs due to walking with large
$\gamma_m$.
To explain this, we denote the second factor as
\begin{eqnarray}
    \sigma_P^{\rm SF}(u,s)
&=& \frac{Z_P^{\rm SF}(L)}{Z_P^{\rm SF}(s L)}
 = \exp\left( \int^{s L}_{L}dL'\frac{\gamma_m^{\rm SF}(u(L'))}{L'}
       \right)
 =  \exp\left( \int_{u}^{\sigma^{\rm SF}(u,s)}du'
               \frac{\gamma_m^{\rm SF}(u')}{\beta^{\rm SF}(u')}
        \right),
 \label{eq:sigP}
\end{eqnarray}
where $u=g^2(L)$ is introduced.
The lower end of the integration range $\sigma^{\rm SF}(u,s)$ is the
solution of
\begin{eqnarray}
 \int_{u}^{\sigma^{\rm SF}(u,s)}\frac{du'}{\beta^{\rm SF}(u')}=\ln(s).
\end{eqnarray}
When $u$ is very close to the fixed point, $\gamma_m$ becomes almost
constant over a wide range of the renormalization scale.
Then, $\sigma_P^{\rm SF}(u,s)$ can be approximated as
\begin{eqnarray}
        \sigma_P^{\rm SF}(u,s)
\approx s^{\gamma_m^*},
\end{eqnarray}
where $\gamma_m^*$ is the mass anomalous dimension at the fixed point.
Substituting $s=M_{\rm ETC}/M_{\rm TC}\sim 1000$ and assuming
$\gamma_m^*\sim 1$, $\sigma_P^{\rm SF}(u,s)$ gives a huge enhancement by
$s$ to the fermion masses in Eq.~(\ref{eq:smf-mass}).

On the lattice, one can calculate $\sigma_P^{\rm SF}(u,s)$.
If the IRFP exists, the mass anomalous dimension at the fixed point is
extracted by
\begin{eqnarray}
   \gamma_m^*
 = \frac{\ln\sigma_P^{\rm SF}(u,s)}{\ln s}.
 \label{eq:gamma_m_star}
\end{eqnarray}

\section{Simulation details}
\label{sec:parameter}

The scale dependence of the gauge coupling and the mass is calculated
in the SF scheme~\cite{Luscher:1992an,Sint:1998iq}.
The detailed setup is almost the same as our previous
work~\cite{Hayakawa:2010yn} except for those subject to the number of
colors, and is described in Appendix~\ref{sec:sf-setup}.
We adopt the unimproved Wilson fermion action and the Wilson
plaquette gauge action, and no improvement is implemented at the action
level.
Instead, at the step of the analysis, the discretization errors are
removed perturbatively as described below.

\subsection{Definition of the running coupling}

With the gauge boundary conditions (\ref{formul:BGF:C0}) and
(\ref{formul:BGF:CT}), the absolute minimum of the action is given by a
color-electric background field denoted by $B(x)$.
Then, the effective action can be defined as a function of $B$ by
\begin{eqnarray}
 \Gamma[B]=-\ln Z_{\rm SF}(C',\bar \rho',\rho'\,; C,\bar \rho,\rho),
\end{eqnarray}
which has the following perturbative expansion in the bare coupling
constant:
\begin{eqnarray}
 \Gamma=\frac{1}{g_0^2}\,\Gamma_0 + \Gamma_1 + O(g_0^2)\,,
\end{eqnarray}
and, in particular, the lowest-order term,
\begin{eqnarray}
 \Gamma_0 = \left[g_0^2\,S_g[B]\right]_{g_0=0},
\end{eqnarray}
is exactly the classical action of the induced background field.
The SF scheme coupling is then defined in the massless limit of
fermions by
\begin{eqnarray}
   \left.\frac{\partial \Gamma}{\partial \eta}
   \right|_{\eta=\pi/4,\,M=0}
 = \frac{1}{g_{\rm SF}^2(g_0^2,\,l)}\,
   \left.\frac{\partial \Gamma_0}{\partial \eta}
   \right|_{\eta=\pi/4,\,M=0}
 = \left. \frac{k}{g_{\rm SF}^2(g_0^2,\,l)}
   \right|_{M=0},
\end{eqnarray}
where $l=L/a$ and $M$ is the mass in the lattice unit defined in the
next subsection.
$\eta=\pi/4$ is chosen following Ref.~\cite{Luscher:1992zx}.
The normalization constant $k$ is determined such that
$g_{\rm SF}^2=g_0^2$ in the leading order of the perturbative expansion,
and is found to be
\begin{eqnarray}
   k
 = \left.\frac{\partial \Gamma_0}{\partial \eta}
   \right|_{\eta=\pi/4,\,M=0}
 = -24\, l^2\, \sin \left[\frac{\pi}{2\,l^2}\right].
\end{eqnarray}
Because of the absence of the $O(a)$ improvement for the fermion action,
only the $\eta$ derivative of the gauge action contributes to $1 /g_{\rm
SF}^2(g_0^2,\,l)$.

\subsection{Definition of $Z_P$}

In the SF setup, the renormalization constant of the pseudoscalar
density and the fermion mass on the lattice are defined by
\begin{eqnarray}
&& \left.
   Z_P^{\rm lat}(g_0^2,L)
 = c \frac{\sqrt{3\,f_1}}{f_P(L/2)}
   \right|_{M=0},
\label{eq:ZP-definition}\\
&& M
 = \left.
     \frac{1}{2}
     \frac{(\partial^* + \partial)f_A(x_0)}
          {2\,f_P(x_0)}\right|_{x_0=L/2}
 \label{eq:pcac-mass},
\end{eqnarray}
respectively, where
\begin{eqnarray}
     f_P(x_0)
 &=& -\frac{1}{N_f^2-1}\sum_{\vec y,\, \vec z}
      \langle
       \overline\psi(x)\gamma_5 T^a \psi(x)\
       \overline\zeta(\vec y)\gamma_5 T^a \zeta(\vec z)
      \rangle,\\
     f_A(x_0)
 &=& -\frac{1}{N_f^2-1}\sum_{\vec y,\, \vec z}
      \langle
       \overline\psi(x)\gamma_\mu \gamma_5 T^a \psi(x)\
       \overline\zeta(\vec y)\gamma_5 T^a \zeta(\vec z)
      \rangle,\\
     f_1
 &=& -\frac{1}{(N_f^2-1)L^6}\sum_{\vec u,\, \vec v,\,\vec y,\, \vec z}
      \langle
       \overline{\zeta'}(\vec u)\gamma_5 T^a \zeta'(\vec v)\
       \overline\zeta(\vec y)\gamma_5 T^a \zeta(\vec z)
      \rangle,
\end{eqnarray}
and $\zeta$ ($\zeta'$) is the boundary fermion at $x^0=0$
($x^0=L$)~\cite{Luscher:1992an}.
$\partial$ and $\partial^*$ are forward and backward lattice derivatives,
respectively, and $T^a$ are the generators of the SU($N_f$) group.
All quantities defined above are dimensionless.
$c$ is determined such that $Z_P^{\rm lat}(g_0^2,L)=1$ at tree level.
We calculate $c$ in the background field method with our setup.
This results in a great advantage in the analysis of the small coupling
region as explained later.
The tree level values of the critical $\kappa$ and $c$ for various
lattice sizes are tabulated in Table~\ref{tab:c_zp}.
%%%%%%%%%%%%%%%%%%%%%%%%%%%%%%%%%%%%%%%%%%%%%%%%%%%%%%%%%%%%%%%%%%%%%%%%%%
\begin{table}[tb]
 \centering
 \begin{tabular}{r|lll}
  $l=L/a$ & $m_c$ & $\kappa_c$ &
  $c=\frac{f_P(L/2)}{\sqrt{3\,f_1}}|_{g_0^2=0,\kappa=\kappa_c}$ \\
  \hline
 6 & -0.0375518783340131  & 0.126184617349584 & 1.2302955268807\\
 8 & -0.0212857986789711  & 0.125668739874301 & 1.35598112469427\\
%10 & -0.0136556191719716  & 0.125428199932927 & 1.42695919467544\\
12 & -0.00949014679324021 & 0.125297272376912 & 1.47058643882059\\
%14 & -0.00697312005024675 & 0.125218290543111 & 1.49928506024377\\
16 & -0.00533796269846268 & 0.125167034239963 & 1.51920100305346\\
18 & -0.00421651481946245 & 0.125131905133095 & 1.53362308991718\\
%20 & -0.00341430454691251 & 0.125106788168924 & 1.5444319707906\\
%22 & -0.00282081630422311 & 0.125088212717475 & 1.55276422045805\\
24 & -0.00236949881524477 & 0.125074090727449 & 1.55933924368169\\
%26 & -0.00201834338917621 & 0.125063105072838 & 1.56463049980813\\
%28 & -0.00173977758770251 & 0.125054391706984 & 1.56896010117542\\
%30 & -0.00151509932798588 & 0.125047364794592 & 1.57255353665288\\
%32 & -0.00133126110570859 & 0.125041615759914 & 1.57557260689087
 \end{tabular}
 \caption{The numerical values of $\kappa_c$ at the tree level and $c$.}
 \label{tab:c_zp}
\end{table}
%%%%%%%%%%%%%%%%%%%%%%%%%%%%%%%%%%%%%%%%%%%%%%%%%%%%%%%%%%%%%%%%%%%%%%%%%%

\subsection{Parameters}

The simulation was performed on the lattice sizes of
$l^4=(L/a)^4$ = $6^4$, $8^4$, $12^4$, $16^4$, $18^4$, and $24^4$ in a
wide range of $\beta=4/g^{2}_{0}$ ($1.7 \le \beta \le 24.0$).

The algorithm to generate the gauge configurations follows the standard
hybrid Monte Carlo (HMC) with three pseudofermion fields and the Omelyan
integrator with $\lambda=0.0708$.
The numerical simulations were carried out on several different
architectures including a general purpose graphics processing unit
and PC cluster.
In order to achieve high performance on each architecture,
the fermion solver part was optimized depending on architecture.
In particular, the mixed precision solver and the flavor-parallelized
blocked HMC algorithm using multiple GPUs enables us to obtain high
statistics~\cite{Hayakawa:2010gm}.
The acceptance rate is kept to around 80 \% by adjusting the
molecular dynamics (MD) step size ($\delta \tau$).
Since the Wilson fermion explicitly breaks chiral symmetry, the value of
$\kappa$ is tuned for every pair of $(\beta,\ L/a)$ to its critical
value $\kappa_c$ by monitoring the quark mass defined in
Eq.~(\ref{eq:pcac-mass}).

\section{Running coupling}
\label{sec:analysis-g2}

\subsection{Numerical results}
\label{sec:results}

The SF coupling constant ($g_{\rm SF}^2$) and the dimensionless quark
mass ($M$) obtained on each ($\beta$, $\kappa$, $l$) are shown in
Tables~\ref{tab:simpara_L6_imp0}-\ref{tab:simpara_L24_imp0} in
Appendix~\ref{sec:raw-data} together with other information such as
the number of accumulated trajectories (Traj.), the MD step size
$\delta \tau$, the acceptance rate (Acc.), and the plaquette value
(plq.).
The data with $|M|$ of typically $O(10^{-4})$ or, at most, 0.003 are
only used in the following analysis.

\begin{figure}[h]
 \centering
 \begin{tabular}{cc}
  \includegraphics*[width=0.5 \textwidth,clip=true]
  {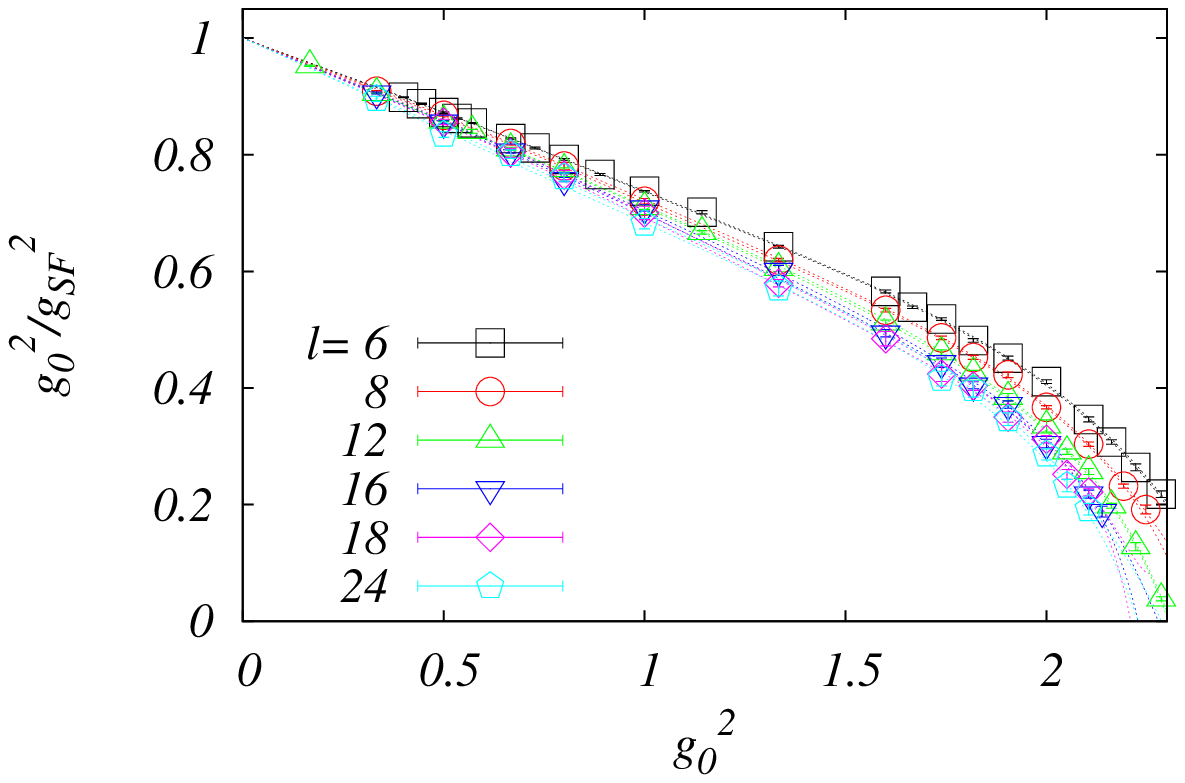}&
  \includegraphics*[width=0.5 \textwidth,clip=true]
  {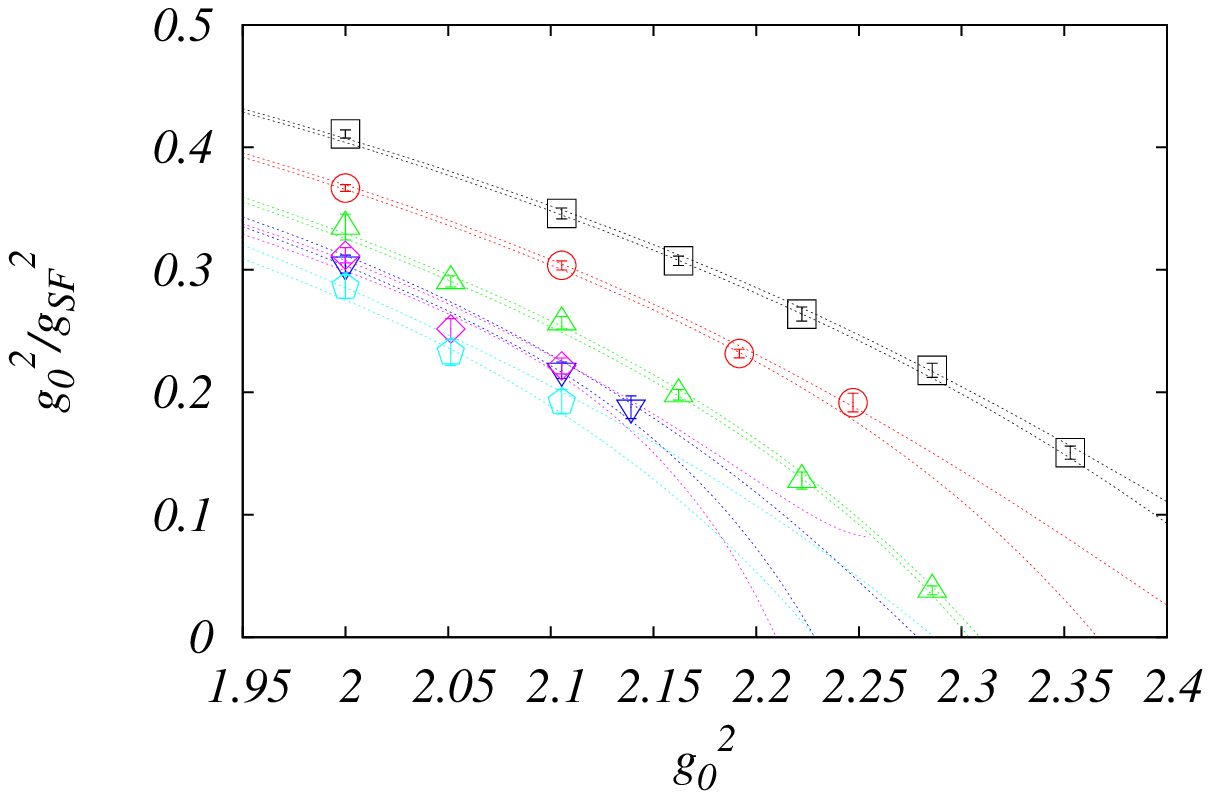}\\
 \end{tabular}
\caption{$g_0^2$ dependence of $g_0^2/g_{\rm SF}^2$.
 The right panel magnifies the region of $g_0^2 \in [1.95,\ 2.40]$.}
\label{fig:beta_vs_g2}
\end{figure}

$g_0^2/g_{\rm SF}^2$ is shown as a function of the bare coupling
constant $g_0^2$ in Fig.~\ref{fig:beta_vs_g2}.
As a general behavior, at a given $g_0^2$, $g_{\rm SF}^2(g_0^2,l)$
increases with $l$, which is consistent with asymptotic freedom.
For later use, the $g_0^2$ dependence of $g_0^2/g_{\rm SF}^2(g_0^2,l)$
is fitted, at each $l$, with the interpolating formula
\begin{eqnarray}
     \frac{g_0^2}{g_{\rm SF}^2(g^{2}_{0}, l)}
 &=& \frac{1-a_{l,1}\,g_0^4}
          {1+p_1(l)\times g_0^2+
           \sum_{n=2}^N a_{l,n} \times g_0^{2\,n}
          },
 \label{eq:fitfunc}
\end{eqnarray}
which is found to be the best among various functional forms we have
tried.
The degree of a polynomial $N$ is varied from 3 to 5 to look for the
best fit.
The coefficients $a_{l,n}$ thus determined are tabulated in
Table~\ref{tab:fit_param1_pade_4_su2}, and the fit results are shown in
Fig.~\ref{fig:beta_vs_g2} as the dotted curves.
%%%%%%%%%%%%%%%%%%%%%%%%%%%%%%%%%%%%%%%
\begin{table}[tbh]
\centering
\begin{tabular}{cc|cccccc}
$l$ & $N$ & $\chi^2$/d.o.f. & $a_{l,1}$ & $a_{l,2}$ & $a_{l,3}$ & $a_{l,4}$ & $a_{l,5}$ \\
\hline
  6 & 3 & 3.3
    &  0.162(1) & $-$0.104( 2) & $-$0.030( 2)\\
  6 & 4 & 1.9
    &  0.167(1) & $-$0.133( 6) &  0.011( 8) & $-$0.017( 3)\\
  6 & 5$^*$ & 0.9
    &  0.162(2) & $-$0.175(13) &  0.142(37) & $-$0.121(29) &  0.026( 7)\\
\hline
  8 & 3 & 2.6
    &  0.172(2) & $-$0.089( 6) & $-$0.042( 5)\\
  8 & 4 & 1.3
    &  0.180(2) & $-$0.137(14) &  0.024(18) & $-$0.026( 7)\\
  8 & 5$^*$ & 1.1
    &  0.176(4) & $-$0.170(30) &  0.133(85) & $-$0.114(66) & 0.022(17)\\
\hline
 12 & 3 & 1.3
    &  0.1873(4)& $-$0.089( 4) & $-$0.055( 3)\\
 12 & 4$^*$ & 0.4
    &  0.1880(4)& $-$0.114(10) & $-$0.011(16) & $-$0.017(6)\\
 12 & 5 & 0.5
    &  0.1881(5)& $-$0.110(20) & $-$0.024(56) & $-$0.006(45) & $-$0.003(11)\\
\hline
 16 & 3 & 1.7
    &  0.189(3) & $-$0.091( 9) & $-$0.051(8)\\
 16 & 4$^*$ & 1.1
    &  0.198(4) & $-$0.148(23) &  0.040(35) & $-$0.040(15)\\
 16 & 5 & 1.2
    &  0.196(7) & $-$0.168(47) &  0.110(141)& $-$0.100(116) &  0.017(31)\\
\hline
 18 & 3 & 1.6
    &  0.184(6) & $-$0.073(17) & $-$0.051(17)\\
 18 & 4$^*$ & 0.9
    &  0.201(6) & $-$0.190(46) &  0.112(62) & $-$0.067(23)\\
 18 & 5 & 1.0
    &  0.201(9) & $-$0.192(108)&  0.117(291)& $-$0.071(227) &  0.001(57)\\
\hline
 24 & 3$^*$ & 1.3
    &  0.197(5) & $-$0.054(17) & $-$0.079(15)\\
 24 & 4 & 1.5
    &  0.199(8) & $-$0.066(54) & $-$0.060(82) & $-$0.008(34)\\
 24 & 5 & 1.5
    &  0.207(7) &  0.012(94) & $-$0.316(247)&  0.211(195) & $-$0.059(50)\\
\end{tabular}
\caption{The coefficients determined by fitting to
 Eq.~(\ref{eq:fitfunc}).
 $N$ with $^*$ is the one chosen in the following analysis.}
\label{tab:fit_param1_pade_4_su2}
\end{table}

%%%%%%%%%%%%%%%%%%%%%%%%%%%%%%%%%%%%%%%
%%%%%%%%%%%%%%%%%%%%%
%%%%%%%%%%%%%%%%%%%%%

In lattice perturbation theory, we calculated $p_1(l)$ in
Eq.(\ref{eq:fitfunc}) and obtained
\begin{eqnarray}
 p_1(l) = \left\{
 \begin{array}{ll}
%0.24949959 & \mbox{ for } L/a=4\\
 0.26506184 & \mbox{ for } L/a=6\\
 0.27347456 & \mbox{ for } L/a=8\\
%0.27913209 & \mbox{ for } L/a=10\\
 0.28356222 & \mbox{ for } L/a=12\\
%0.28731231 & \mbox{ for } L/a=14\\
 0.29062046 & \mbox{ for } L/a=16\\
 0.29360965 & \mbox{ for } L/a=18\\
%0.29635214 & \mbox{ for } L/a=20\\
%0.29889473 & \mbox{ for } L/a=22\\
 0.30127006 & \mbox{ for } L/a=24.\\
%0.30350217 & \mbox{ for } L/a=26\\
%0.30560956 & \mbox{ for } L/a=28\\
%0.30760690 & \mbox{ for } L/a=30\\
%0.30950612 & \mbox{ for } L/a=32
 \end{array}
 \right.
 \label{eq:p_1_su2}
\end{eqnarray}
These values will be used in the perturbative improvement of the
discretization error.

\subsection{Discrete $\beta$ function}
\label{subsec:DBF}

Hereafter the subscript ``SF'' is omitted as no confusion will arise.
Since the raw data of $1/g^2(g_0^2, l)$ fluctuate around zero in the
strong coupling region, converting to $g^2(g_0^2,l)$ sometimes induces
a huge statistical uncertainty.
To avoid this, we deal with the inverse coupling constant
$1/g^2(g_0^2,l)$ in the analysis.
The running of the coupling is analyzed with the discrete $\beta$
function (DBF)~\cite{Shamir:2008pb}
\begin{eqnarray}
     B(u,s)
 &=& \frac{1}{\sigma(u,s)} - \frac{1}{u},
\end{eqnarray}
where
\begin{eqnarray}
     \sigma(u,s)
 &=& g^2(sL)|_{u=g^2(L)},
\label{eq:dbf-cont}
\end{eqnarray}
where $s> 1$ denotes the change of the renormalization scale.
In this work, we take $s$=3/2 or 2.

At the leading order (LO) of continuum perturbation theory, the DBF is
constant,
\begin{eqnarray}
     B^{\rm LO}(u,s)
 &=& -b_1\ln(s)
  = \left\{\begin{array}{cl}
%    -0.012145120 & \mbox{for $s=4/3$}\\
     -0.017117585 & \mbox{for $s=3/2$}\\
     -0.029262705 & \mbox{for $s=2$}.
           \end{array}
    \right.
\end{eqnarray}
The calculation including higher order terms is straightforward.
Later, the nonperturbative results are compared with the two-loop
perturbative result.

On the lattice, we define the lattice DBF in the same manner by
\begin{eqnarray}
     B^{\rm lat}(u,s,l)
 &=& \frac{1}{\Sigma(u,s,l)}
   - \frac{1}{u},\\
     \Sigma(u,s,l)
 &=& g^2(g_0^2,s\cdot l)|_{u=g^2(g_0^2,l)}.
\label{eq:dbf}
\end{eqnarray}

The scale dependence of the coupling in the continuum limit is
extracted by the step scaling technique as follows.
Here, $g^2$ and $g_0^2$ denote the continuum and the lattice bare
coupling, respectively.
First, we choose an initial value of the renormalized coupling constant,
denoted by $u$, which implicitly sets the initial length scale $L$
through $u=g^2(L)$.
Using the interpolating formula, Eq.~(\ref{eq:fitfunc}), at each lattice
size $l$, the corresponding bare coupling constant ${g_0}^2$ is
numerically determined by solving the equation $u=g^2({g_0}^2,\,l)$.
The lattice step scaling function $\Sigma(u,s,l)$ is then defined as
the SF coupling at the length scale $s\cdot l$ and the same bare
coupling ${g_0}^2$, Eq.~(\ref{eq:dbf}).
Since both $l$ and $s \cdot l$ must be equal to one of 6, 8, 12, 16, 18,
and 24, the possible values for the rescaling factor $s$ are limited.
The difference between $\Sigma(u,s,l)$ and $u$ gives the scale
dependence up to lattice artifacts.

By repeating the same procedure at a fixed $u$ but with different $l$
and taking the continuum limit, the lattice artifacts can be removed.
We calculate the continuum limit of this function for various initial
values $u$.
In asymptotic-free theories, the DBF is negative at a small coupling
region.
If the sign of the continuum DBF flips at a certain renormalized
coupling constant $u$, it indicates the existence of the IRFP around
there.

\subsection{Improving discretization errors}
\label{subsec:improving-dbf}

Since we employ unimproved lattice actions, our results may be contaminated by
substantial $O(a)$ discretization errors.
In principle, those errors can be removed by the continuum
extrapolation, but it may require unreasonably large
lattices as we will see below~\cite{Sint:2011gv}.
Thus, with limited resources, it is important to remove discretization errors
as much as possible before taking the continuum limit.
To do this, we perform the perturbative improvement on the step scaling
function as follows.

In continuum perturbation theory, the step scaling function
$\sigma(u,s)$ is given by
\begin{eqnarray}
&&  \sigma(u,s) = u + s_0 u^2 + s_1 u^3 + s_2 u^4 + \cdots,
       \label{eq:sigma}\\
&& s_0 = {b_1} \ln (s),\\
&& s_1 = \left({b_1}^2 \ln (s)+{b_2}\right)\ln (s) ,\\
&& s_2 = \left(   {b_1}^3 \ln^2(s)
                + \frac{5}{2} {b_1} {b_2} \ln(s)
                + {b_3}
         \right)\ln(s),
\end{eqnarray}
where $b_i$'s are the coefficients of the $\beta$ function introduced in
Eq.~(\ref{eq:betafunc}).
Let $B_0^{\rm lat}(u,s,l)$ and $\Sigma_0(u,s,l)$ be the unimproved
lattice DBF and the step scaling function, respectively.
Then the difference between the continuum and lattice DBF is
\begin{eqnarray}
    B(u,s)-B_0^{\rm lat}(u,s,l)
&=&
    \frac{\Sigma_0(u,s,l)-\sigma(u,s)}{\sigma(u,s)\Sigma_0(u,s,l)}
 =  \frac{\delta_0(u,s,l)}{\Sigma_0(u,s,l)},
 \label{eq:DBF-difference}
\end{eqnarray}
where we have introduced the measure of the discretization error as
\begin{eqnarray}
   \delta_0(u,s,l)
 = \frac{\Sigma_0(u,s,l)-\sigma(u,s)}
        {\sigma(u,s)}
  = \delta^{(1)}(s,l)\, u + O(u^2).
 \label{eq:Sig-sig-0}
\end{eqnarray}

With $p_1(l)$ given in Eq.~(\ref{eq:p_1_su2}), the coefficient of the
leading order term $\delta^{(1)}(s,l)$ is calculated as
\begin{eqnarray}
   \delta^{(1)}(s,l)
 = \Big( p_1(s\,l)-b_{1}\ln(s\,l) \Big)
 - \Big( p_1(   l)-b_{1}\ln(   l) \Big)
 = p_1(s\,l)-p_1(l)-b_1\ln(s).
 \label{eq:delta_1}
\end{eqnarray}
$\delta^{(1)}(s,l)$ is tabulated in Table~\ref{tab:delta2}.
It is seen that, for a fixed $s$, the change of $\delta^{(1)}(s,l)$ with
$l$ is not monotonic.
The same is observed in the improvement coefficient in ten-flavor
QCD~\cite{Hayakawa:2010yn}.
This nonmonotonic behavior indicates that when using unimproved actions
without any improvement the continuum limit gives a wrong value unless
$l$ is extremely large~\cite{Sint:2011gv}.
Thus, the one-loop improvement is important.
The same may happen to the two-loop correction, which is not available.
But, since the coefficients of one-loop correction are reasonably small,
we expect the effect is small at the two-loop level or higher.
%%%%%%%%%%%%%%%%%%%%%%%
\begin{table}
 \centering
 \begin{tabular}{c|ccc}
  $(s,l)$ & $(3/2, 8)$ & $(3/2,12)$ & $(3/2,16)$\\
  $\delta^{(1)}(s,l)$ & $-0.007030$ & $-0.007070$ & $-0.006468$\\
\hline
  $(s,l)$ & $(  2, 6)$ & $(  2, 8)$ & $(  2,12)$ \\
  $\delta^{(1)}(s,l)$ & $-0.010762$ & $-0.012117$ & $-0.011555$\\
 \end{tabular}
 \caption{Coefficients for perturbative correction, $\delta^{(1)}(s,l)$,
 for $(s,l)$.
 }
 \label{tab:delta2}
\end{table}
%%%%%%%%%%%%%%%%%%%%%%%

Replacing $\Sigma_0(u,s,l)$ in Eqs.~(\ref{eq:DBF-difference}) and
(\ref{eq:Sig-sig-0}) with the one-loop improved one,
\begin{eqnarray}
\Sigma_1(u,s,l)=\frac{\Sigma_0(u,s,l)}{1+\delta^{(1)}(s,l)\,u}, 
\end{eqnarray}
the discretization error reduces to $O(u^2)$.
In the following, we mainly analyze the one-loop improved DBF defined by
\begin{eqnarray}
     B_1^{\rm lat}(u,s,l)
 &=& \frac{1}{\Sigma_1(u,s,l)} - \frac{1}{u}.
 \label{eq:dbf-1}
\end{eqnarray}

It should be noted that we have removed the $O(u)$ discretization
error but not the whole $O(a)$ error.
To be precise, the remaining discretization error of the lattice DBF,
{\it i.e.}, $B(u,s)- B_1^{\rm lat}(u,s,l)$, has a form of asymptotic
expansion in $1/l$~\cite{Bode:1999sm} as
\begin{eqnarray}
    B_1^{\rm lat}(u,s,l) - B(u,s)
&=& \left(\frac{1}{l}-\frac{1}{s\,l}\right) e(u) + O(l^{-2}),
\label{eq:error-i}
\end{eqnarray}
where $e(u)$ is a coefficient associated with the leading discretization
error and is an asymptotic series of $u$.
After the one-loop improvement, $e(u)$ is of $O(u^2)$.
Thus, the leading discretization error is still $O(a)$ with a reduced
coefficient, and the extrapolation to the continuum limit will be
performed linearly in $a$.

\subsection{Extraction of the continuum DBF}
\label{subsec:contlim}

The continuum limit is taken for a fixed rescaling factor $s$=3/2 or 2
and with an input value of $u$.
The extrapolation is carried out for every jackknife ensemble, and the
statistical error in the continuum limit is estimated by the single
elimination jackknife method.

%%%%%%%%%%%%%%%%%%%%%
\begin{figure}[tb]
\centering
\begin{tabular}{cc}
\includegraphics*[width=0.5 \textwidth,clip=true]
{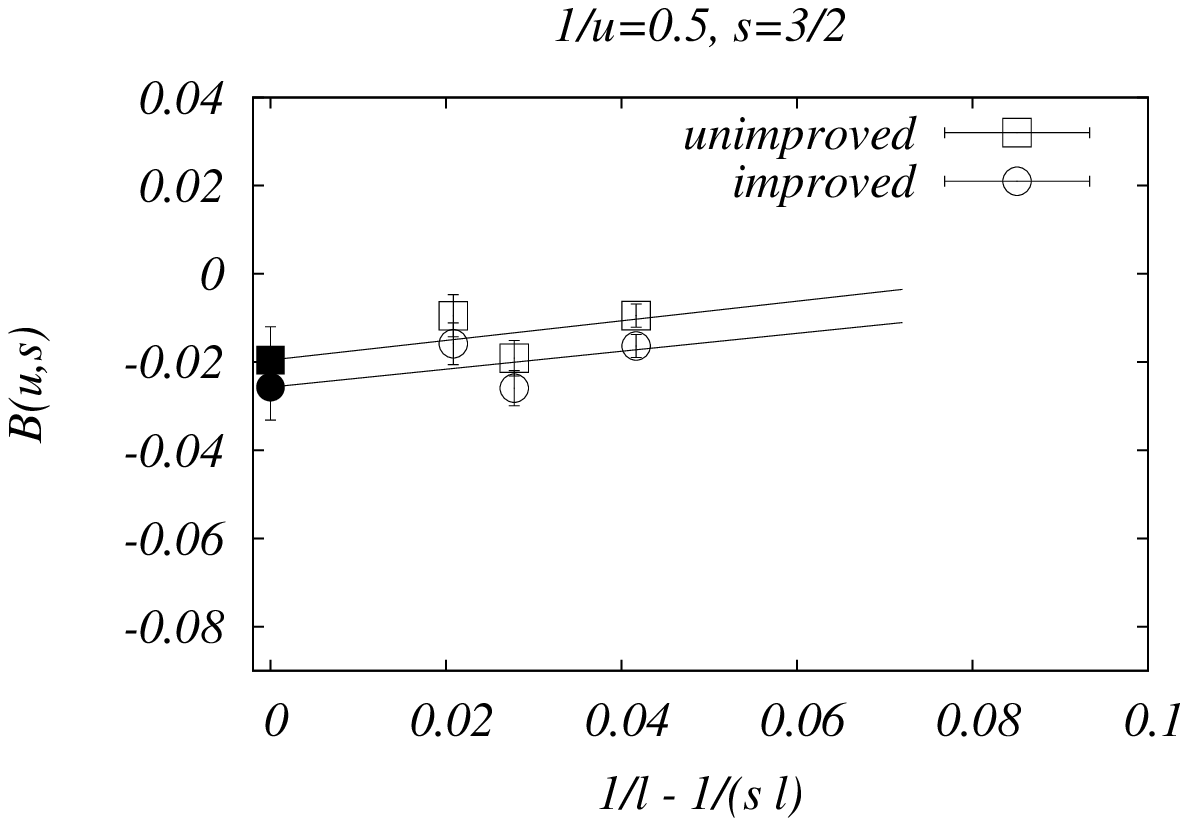}&
\includegraphics*[width=0.5 \textwidth,clip=true]
{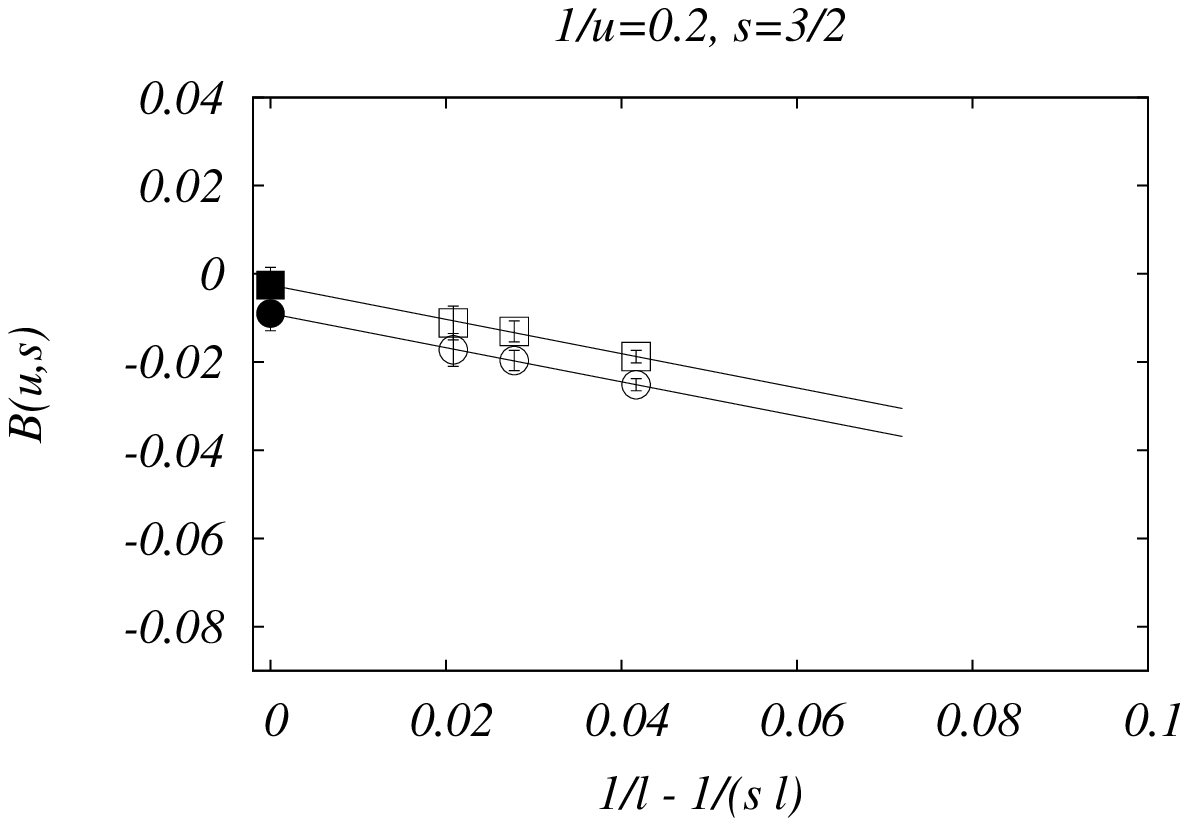}\\ 
\includegraphics*[width=0.5 \textwidth,clip=true]
{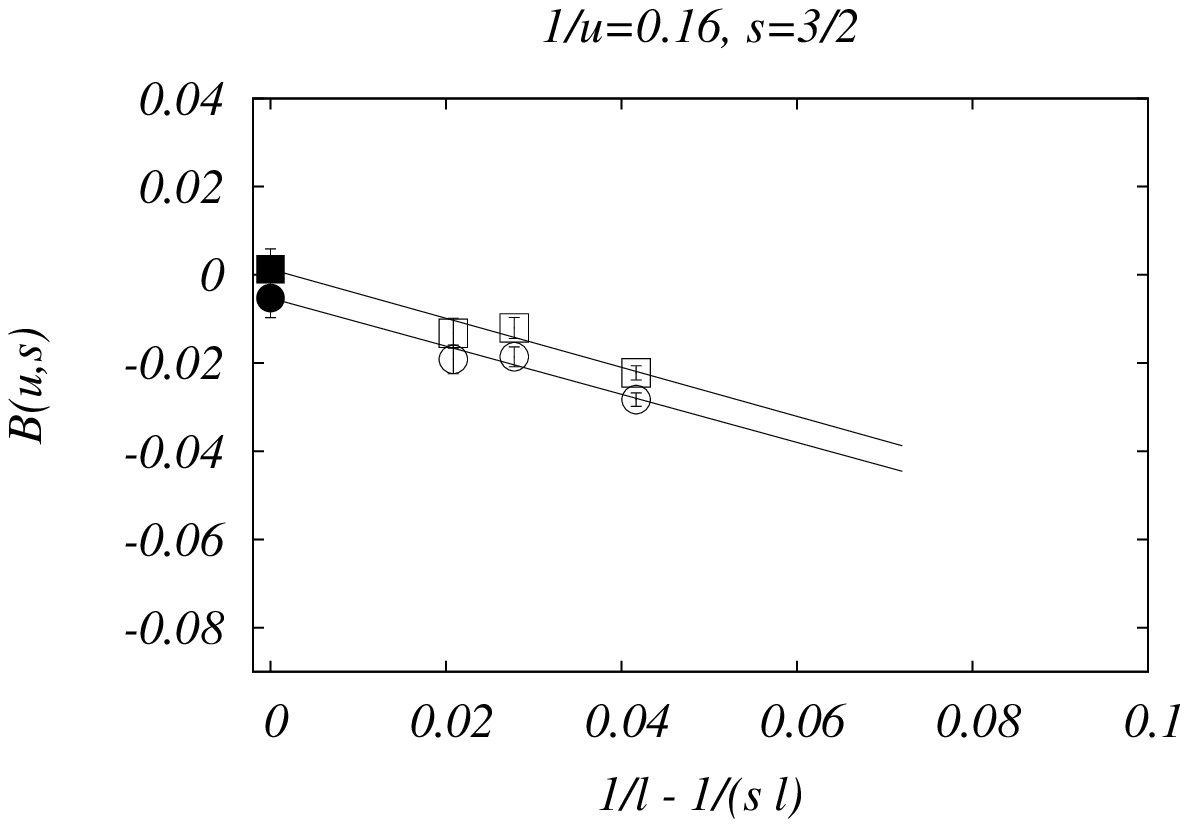}&
\includegraphics*[width=0.5 \textwidth,clip=true]
{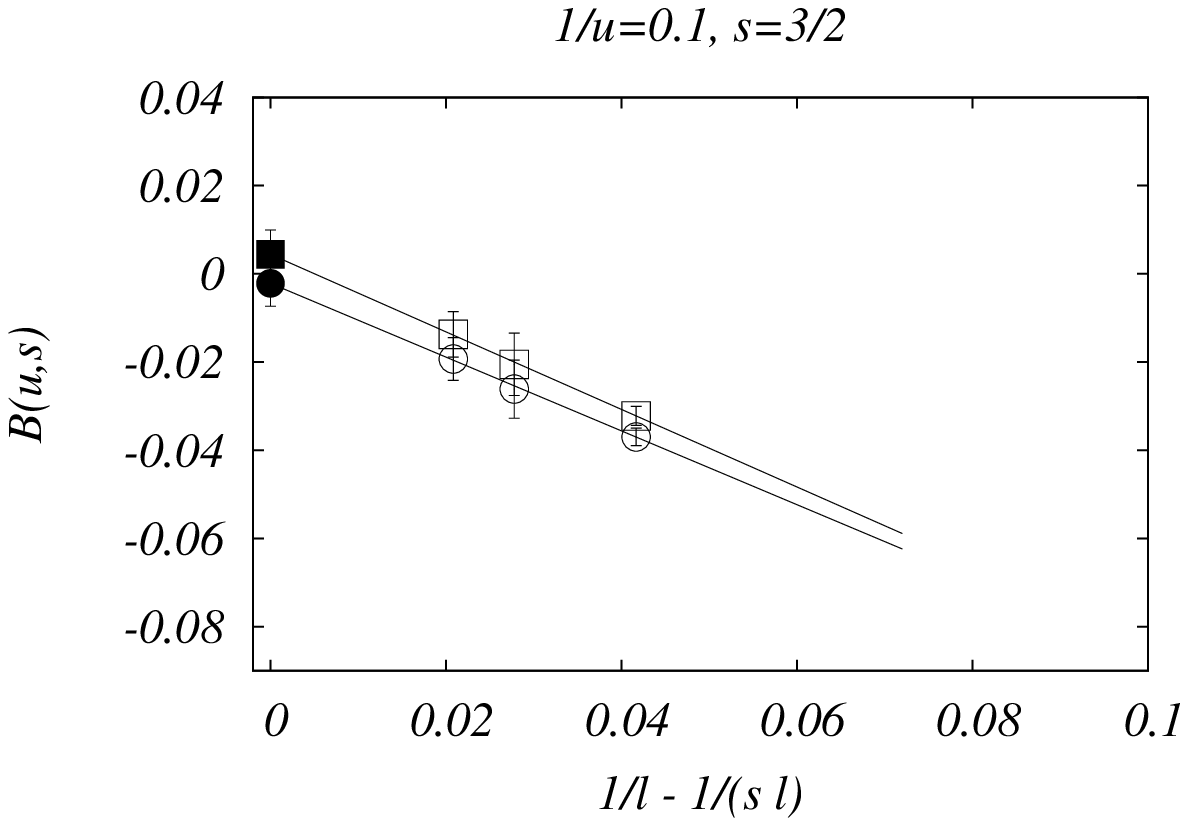}\\
\end{tabular}
\caption{
 Linear extrapolation of DBF to the continuum limit for $s=1.5$.
 }
\label{fig:contlimit-g2run-each-s1.5}
\end{figure}
%%%%%%%%%%%%%%%%%%%%%%%
%%%%%%%%%%%%%%%%%%%%%
\begin{figure}[tb]
\centering
\begin{tabular}{cc}
\includegraphics*[width=0.5 \textwidth,clip=true]
{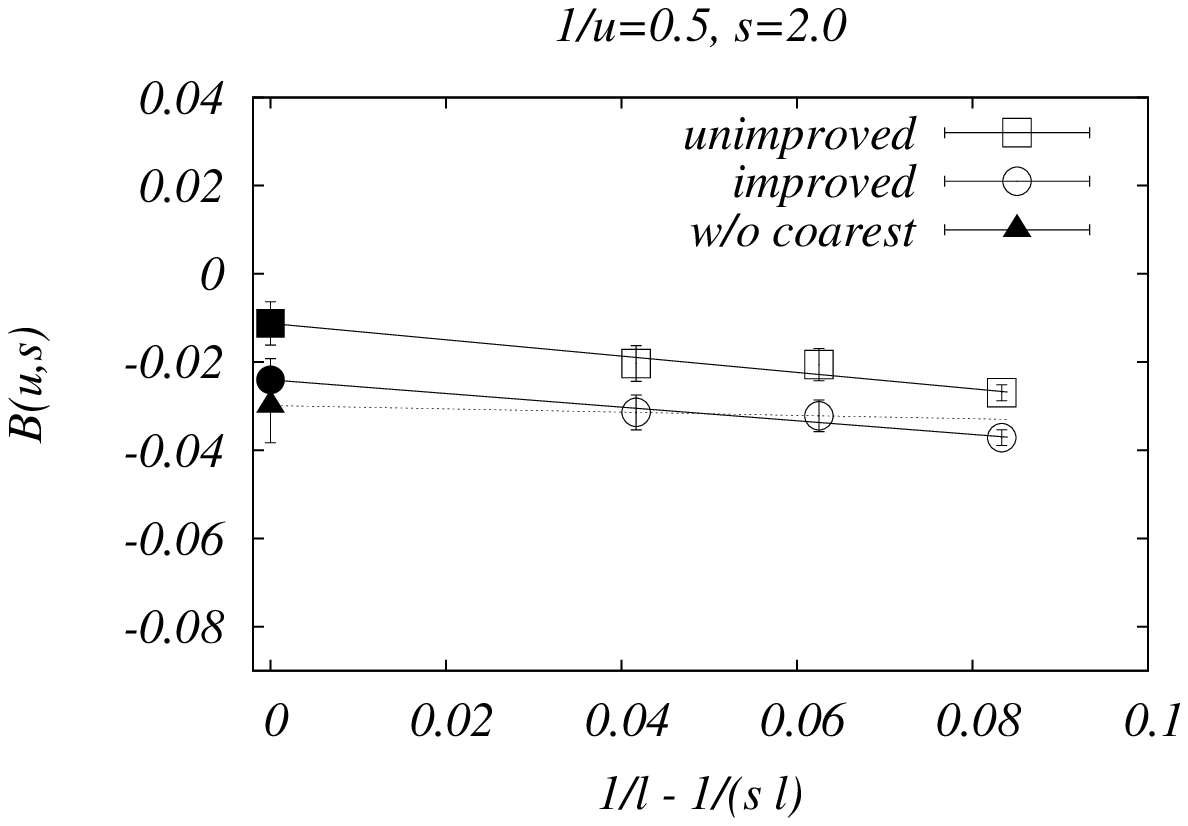}&
\includegraphics*[width=0.5 \textwidth,clip=true]
{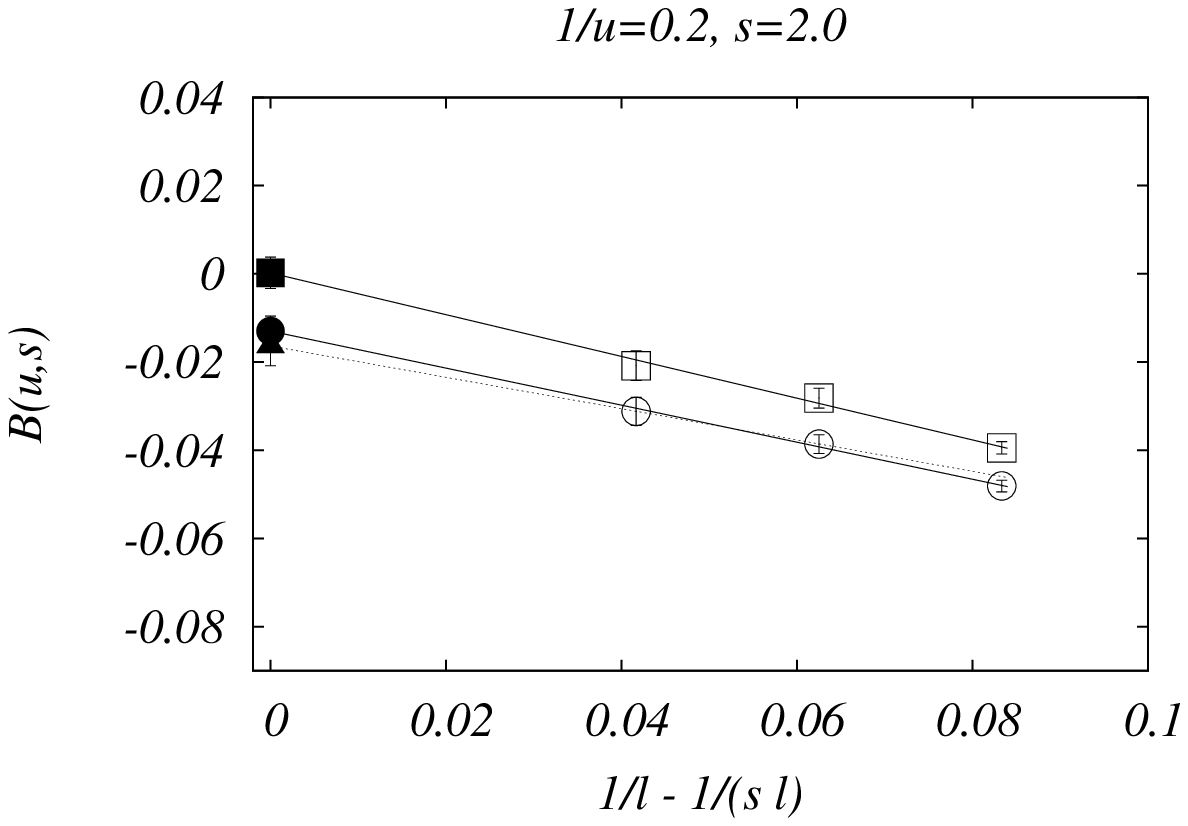}\\ 
\includegraphics*[width=0.5 \textwidth,clip=true]
{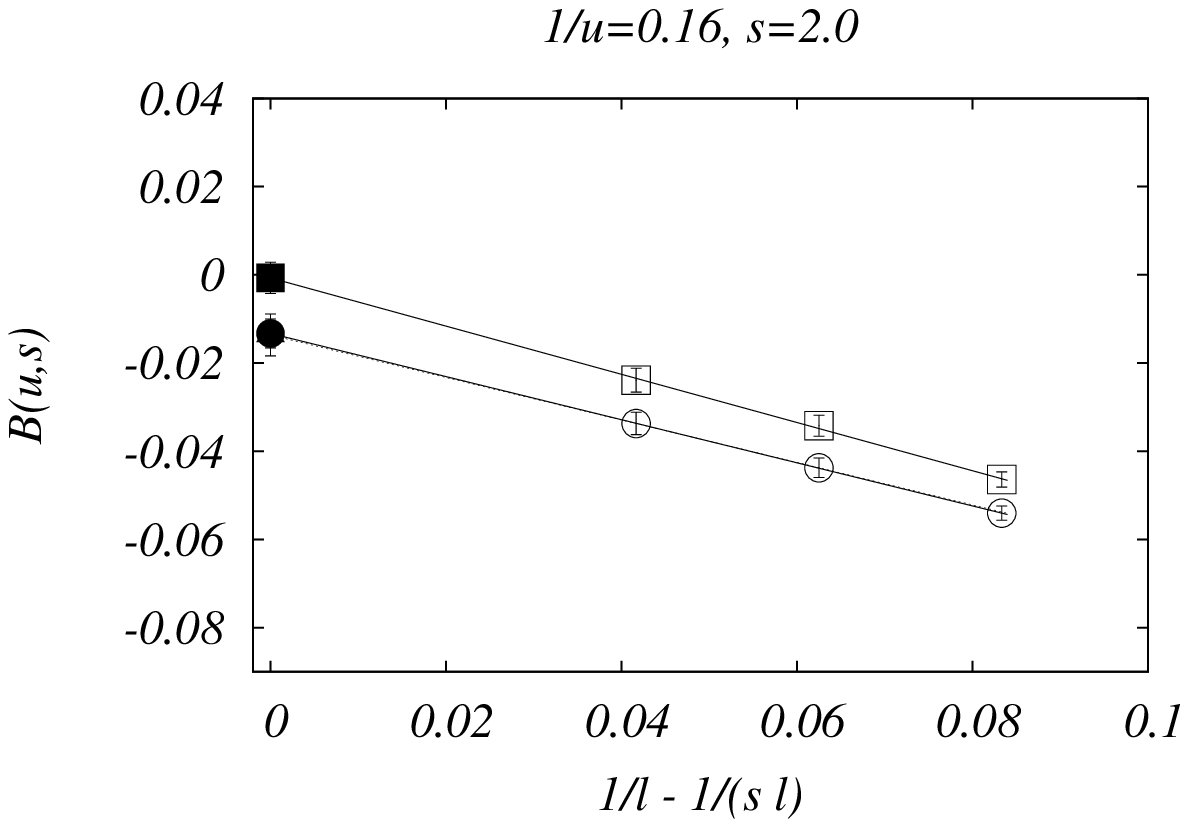}&
\includegraphics*[width=0.5 \textwidth,clip=true]
{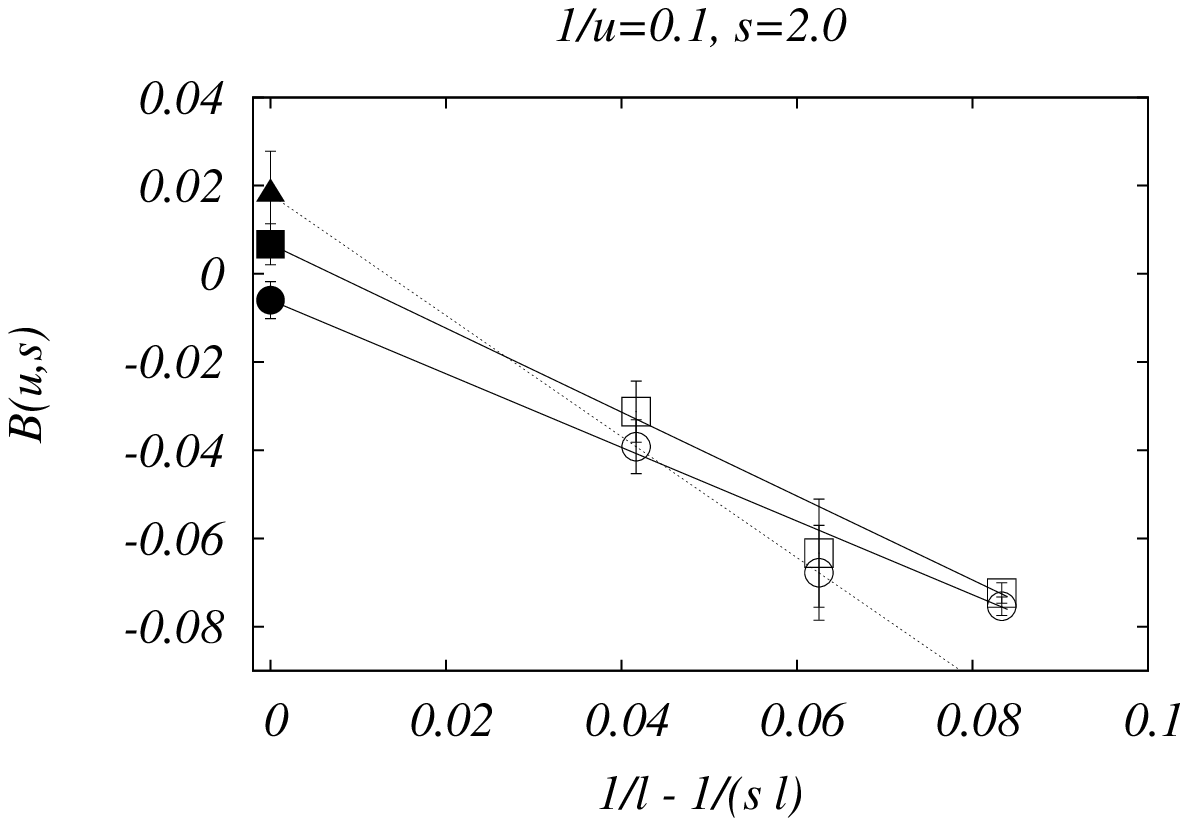}\\
\end{tabular}
\caption{
 Linear extrapolation of DBF to the continuum limit for $s=2$.
 }
\label{fig:contlimit-g2run-each-s2.0}
\end{figure}
%%%%%%%%%%%%%%%%%%%%%%%
Figures~\ref{fig:contlimit-g2run-each-s1.5} and
\ref{fig:contlimit-g2run-each-s2.0} show the continuum limit of
$B_1^{\rm lat}(u,s,l)$ for $s=1.5$ and 2 at the four representative
values of $1/u$ corresponding to $u=2$ - 10, where the values in the
continuum limit are shown in filled symbols.
The results with (circles) and without (squares) perturbative
improvement are shown there.
It turns out that the improvement shifts the data and the continuum
limit downward, and thus the continuum values with and without
improvement disagree as expected from the nonmonotonic behavior of
$\delta^{(1)}(s,l)$.

While at $1/u = 0.5$ the data show a small scaling violation or
even no violation, the nonzero slope clearly appears in the strong
coupling region.
In general, the linear extrapolation appears to be valid, and hence the
continuum limit is expected to be reliable.
However, at $1/u=0.1$ and $s=2$, the data do not align, although the
quality of the linear fit is still acceptable.
A possible reason for this is that the step scaling function at the
coarsest point for $s=2$ contains the $l=6$ data and the $O(a^2)$
discretization error becomes sizable at $1/u=0.1$.
To get rid of the possible contamination due to $O(a^2)$ errors, we fit
the data without the coarsest point at $s=2$; the results of which are
shown in Fig.~\ref{fig:contlimit-g2run-each-s2.0} as dotted lines and
the filled triangles.
It is seen that both fits agree well with each other except at
$1/u=0.1$.
Importantly, the sign of the continuum DBF flips between $1/u$=0.16
and 0.10, indicating the existence of the IRFP somewhere in this region.
Since there is no reason to stick to using the full data, we adopt the
result of the latter analysis as our central value for $s=2$, and the
difference between two analyses is taken into account as the systematic
uncertainty.

On the other hand, the data for $s=3/2$ do not contain the $l=6$ data,
and indeed even at $1/u=0.1$ the data well align.
Thu,s we do not omit the coarsest point for $s=3/2$.

%%%%%%%%%%%%%%%%%%%%%%%
\begin{figure}[tb]
\centering
\begin{tabular}{cc}
\includegraphics*[width=0.53 \textwidth,clip=true]
{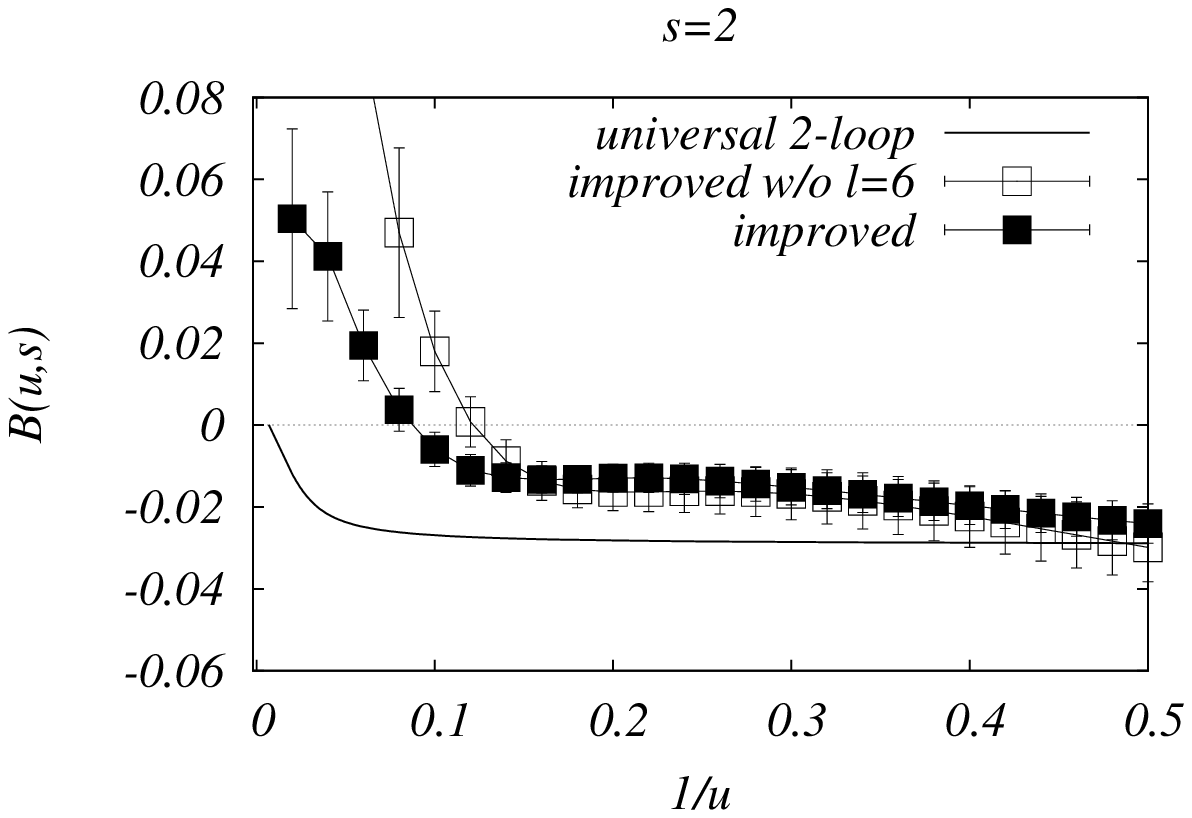}&
\includegraphics*[width=0.53 \textwidth,clip=true]
{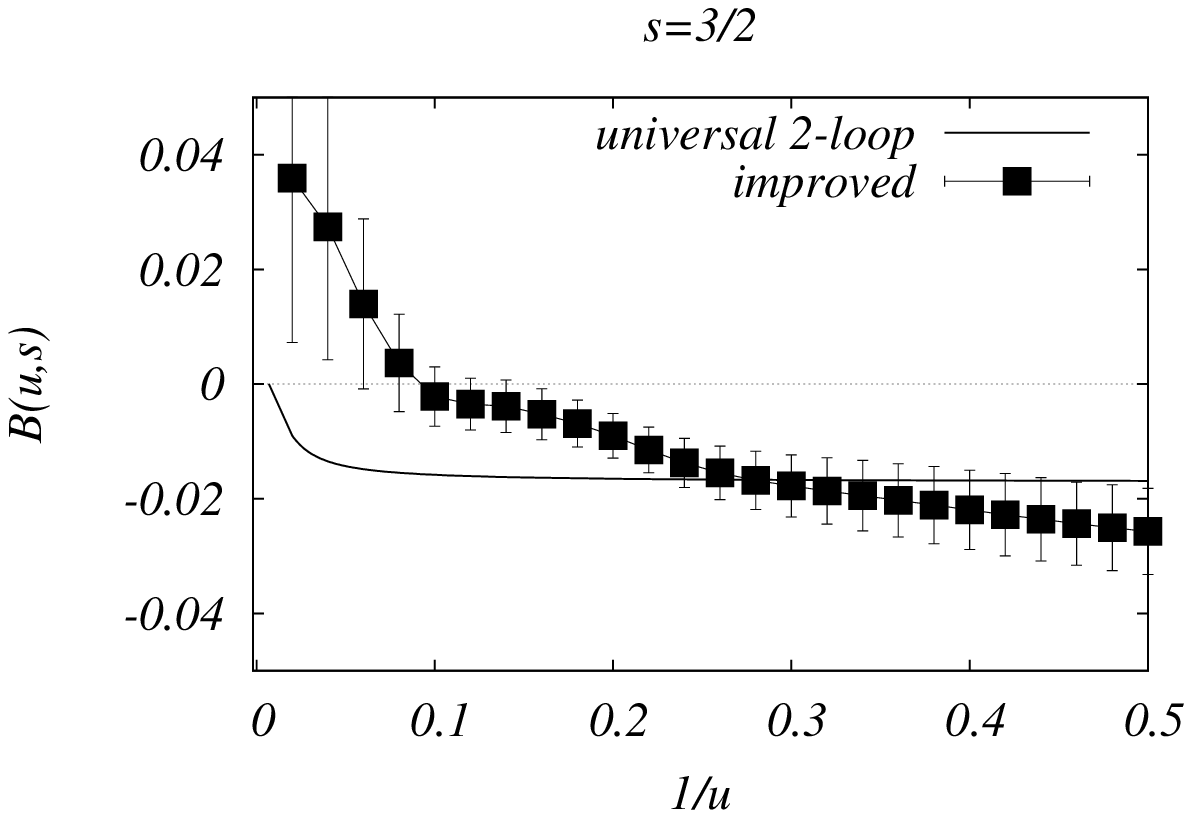}\\
\end{tabular}
\caption{
 $1/u$ dependence of $B(u,s)$ with $s$=2 (left) and 1.5 (right).
 The universal two-loop perturbative prediction is also shown.}
\label{fig:udep-dbf}
\end{figure}
%%%%%%%%%%%%%%%%%%%%%%%
The DBF with one value of $s$ does not have to agree with that with a
different value of $s$.
However, if the IRFP exists, the DBF for arbitrary $s$ vanishes at
the IRFP.
Figure \ref{fig:udep-dbf} shows the $1/u$ dependence of the continuum
DBF with $s=$2 (left) and $s=3/2$ (right), where the results are
compared with the perturbative result at the two-loop approximation.

The continuum DBF without the coarsest data point for $s$=2  (open
squares) is consistent with zero in $0.11 \le 1/u \le 0.13$, which means
that in this region the running coupling constant reaches an IRFP.
Note that, in the region of $u$ where the DBF is positive it is
nontrivial for the continuum limit to exist.
It is observed that the continuum limit using the full data set for
$s=2$ appears to reach IRFP at $1/u\sim 0.1$ smaller than the case
without the coarsest data point.

The behavior of the continuum DBF with $s=3/2$ (right panel of
Fig.~\ref{fig:udep-dbf}) is similar to that with $s=2$.
In this case, the possible location of the IRFP is slightly more
ambiguous than the $s=2$ case due to a larger statistical error.
We observe that the continuum DBF is consistent with zero
in $0.06\le 1/u \le 0.15$ for $s=3/2$.
Since this uncertainty covers the possible range of the IRFP for $s$=2
with and without the coarsest ($l$=6) data point, we take the results at
$s$=3/2 as the conservative estimate for the IRFP.

\section{Running mass}
\label{sec:analysis-mass}

\subsection{Numerical results of $Z_P^{\rm lat}$}

\begin{figure}[tb]
\centering
\begin{tabular}{c}
\includegraphics*[width=0.7 \textwidth,clip=true]
{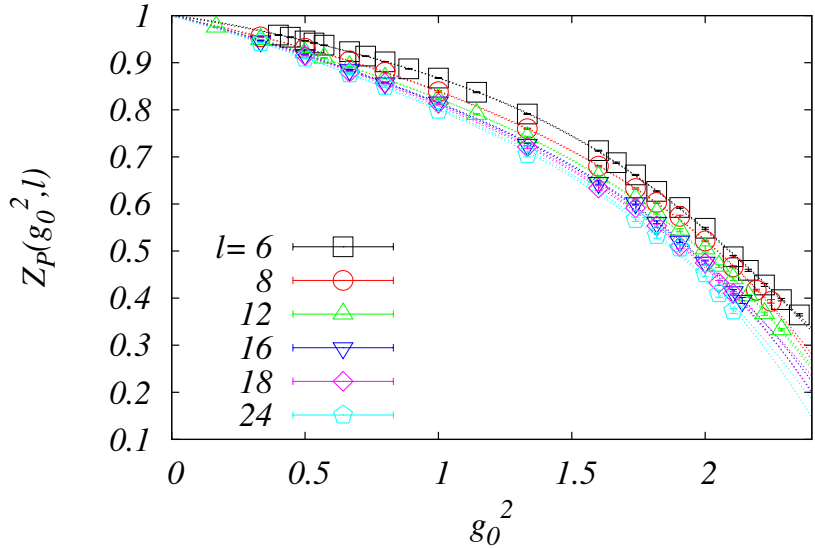}
\end{tabular}
\caption{$Z_P^{\rm lat}(g_0^2,l)$}
\label{fig:zp_lat}
\end{figure}

$Z_P^{\rm lat}(g_0^2,\,l)$ is calculated for various $g_0^2$ and $l$,
and the data are fitted by
\begin{eqnarray}
 Z_P^{\rm lat}(g_0^2,l) = 1 + \sum_{i=1}^N z_i (g_0^2)^i,
 \label{eq:zp-fitfunc}
\end{eqnarray}
for later use.
Fit results are shown in Table~\ref{tab:fit_param1_pade_4_su2_zp} and
Fig.~\ref{fig:zp_lat}.
The degree of the polynomial in Eq.~(\ref{eq:zp-fitfunc}) is chosen to
be the minimal value yielding $\chi^2/$d.o.f. $\le$2.
%%%%%%%%%%%%%%%%%%%%%%%%%%%%%%%%%%%%%%%%%%%%%%%%%%%%%%%%%%%%%%%%%%%%%%%%%%%
%\begin{landscape}
\begin{table}[htb]
\centering
\begin{tabular}{cc|cccccc}
$L/a$ & $N$ & $\chi^2$/d.o.f. & $z_{L/a,1}$ & $z_{L/a,2}$ & $z_{L/a,3}$ & $z_{L/a,4}$ & $z_{L/a,5}$
 %& $z_{L/a,6}$
\\
\hline
  6 &  3  & 3.8 & -0.0974(10) & -0.0057(19) & -0.0293(7)\\
  6 &  4  & 3.5 & -0.0913(31) & -0.0234(87) & -0.0149(68) & -0.0035(16)\\
  6 &  5$^*$  & 2.0 & -0.066(11)  & -0.125(41)  &  0.122(54)  & -0.077(29)
          &  0.0136(52)\\
%  6 &  6  & 1.0 & -0.109(19) &  0.091(92)  & -0.28(17)   &  0.26(14)
%          & -0.118(53) &  0.0193(77)\\
\hline
  8 &  3  & 4.8 & -0.1370(19) &  0.0110(32) & -0.0313(12)\\
  8 &  4  & 2.4 & -0.1197(47) & -0.045(14)  &  0.016(12)  & -0.0116(28)\\
  8 &  5$^*$  & 1.9 & -0.1365(90) &  0.029(37)  & -0.089(49)  &  0.046(26) & -0.0109(48)\\
\hline
 12 &  3  & 7.4 & -0.1517(17) &  0.0051(41) & -0.0284(16)\\
 12 &  4$^*$  & 1.4 & -0.1410(23) & -0.0392(86) &  0.0166(82) & -0.0124(22)\\
% 12 &  5  & 1.4 & -0.1433(58) & -0.026(31)  & -0.006(49)  &  0.002(30) & -0.0029(59)\\
\hline
 16 &  3  & 3.8 & -0.1624(34) &  0.0065(67) & -0.0281(27)\\
 16 &  4$^*$  & 1.4 & -0.1398(64) & -0.069(19)  &  0.043(17)  & -0.0195(46)\\
% 16 &  5  & 1.3 & -0.152(14)  & -0.013(62)  & -0.042(89)  &  0.030(51)  & -0.010(10)\\
\hline
 18 &  3  & 2.0 & -0.1714(52) &  0.0147(93) & -0.0311(35)\\
 18 &  4$^*$  & 1.8 & -0.148(15)  & -0.052(42)  &  0.025(34)  & -0.0141(84)\\
% 18 &  5  & 2.0 & -0.169(39)  &  0.03(14)   & -0.08(19)   &  0.04(10) & -0.011(19)\\
\hline
 24 &  3  & 2.6 & -0.1743(51) &  0.007(10) & -0.0295(40)\\
 24 &  4$^*$  & 1.6 & -0.1529(92) & -0.070(29) &  0.044(27)  & -0.0204(73)\\
% 24 &  5  & 1.3 & -0.189(23)  &  0.11(11)  & -0.23(16)   &  0.144(94) & -0.033(19)\\
\end{tabular}
\caption{The coefficients determined in the fit of $Z_P^{\rm
 lat}(g_0^2,l)$.
 $N$ with $^*$ is the one chosen in the following analysis.}
\label{tab:fit_param1_pade_4_su2_zp}
\end{table}

%\end{landscape}
%%%%%%%%%%%%%%%%%%%%%%%%%%%%%%%%%%%%%%%%%%%%%%%%%%%%%%%%%%%%%%%%%%%%%%%%%%%
%%%%%%%%%%%%%%%%%%%%%%%
%%%%%%%%%%%%%%%%%%%%%%%

Since the statistical errors of $Z_P^{\rm lat}$ are smaller than those
of the coupling constant, the coefficients are well determined.
One of the advantages in our calculation is that $Z_P^{\rm lat}$ defined
in Eq.~(\ref{eq:ZP-definition}) includes a factor $c$, which we have
explicitly calculated as shown in Table~\ref{tab:c_zp}.
Because of this, the first term in the rhs of
Eq.~(\ref{eq:zp-fitfunc}) can be set to unity, which makes the fit
stable.
If $c$ were not available, the first term would be replaced with another
free parameter, making the statistical error somewhat larger.
In Fig.~\ref{fig:zp_lat}, it is seen that $Z_P^{\rm lat}$ decreases with
$l$ at a fixed $g_0^2$.
It is the tendency which is necessary for WTC to work.

\subsection{Step scaling function and its improvement}

The lattice step scaling function for $Z_P$ is defined by
\begin{eqnarray}
  \Sigma_{P,0}^{\rm lat}(u,s,l)
= \left.
    \frac{Z_P^{\rm lat}(g_0^2,\,l)}
         {Z_P^{\rm lat}(g_0^2,\,s\cdot l)}\right|_{u=g^2_{\rm SF}(g_0^2,l)}.
\end{eqnarray}
We implement the one-loop improvement as for the running coupling.
To this purpose, we define the measure of the discretization error by
\begin{eqnarray}
     \delta_{P,0}(u,s,l)
 &=& \frac{\Sigma_{P,0}^{\rm lat}(u,s,l)-\sigma_P^{\rm PT}(u,s)}
          {\sigma_P^{\rm PT}(u,s)}
  =  \delta_P^{(1)}(s,l) u + O(u^2),
 \label{eq:delta_P}
\end{eqnarray}
In principle, the improvement coefficient $\delta_P^{(1)}(s,l)$ can be
calculated by perturbation theory.
Since the coefficient respecting our SF setup is not available,
we follow the prescription adopted in Ref.~\cite{Aoki:2010wm} and
determine it by fitting $\delta_{P,0}$ to a linear function of $u$.
In the fit to determine $\delta_P^{(1)}(s,l)$, $\sigma_P^{\rm PT}(u,s)$
in Eq.~(\ref{eq:delta_P}) has to be specified.
We take the one-loop prediction
\begin{eqnarray}
     \sigma_P(u,s)|_{\rm LO}
 &=& \left(\frac{\sigma(u,s)}{u}\right)^{\frac{27}{40}},
\label{eq:sigP-LO}
\end{eqnarray}
as $\sigma_P^{\rm PT}(u,s)$.
The fit results are shown in Table~\ref{tab:deltaP}.
It is seen that the coefficient is small except for that with $l$=6.
%%%%%%%%%%%%%%%%%%%%%%%
\begin{table}[tb]
 \centering
 \begin{tabular}{c|ccc}
  $(s,l)$ & (3/2, 8) & (3/2,12) & (3/2,16)\\
  $\delta_P^{(1)}(s,l)$ & 0.0051(7) & $-$0.0001(13) & 0.0001(14)\\
\hline
  $(s,l)$ & (2,  6) & (2,  8) & (2, 12)\\
  $\delta_P^{(1)}(s,l)$ & 0.0216(5) & 0.0040(9) & $-$0.0002(13)\\
 \end{tabular}
 \caption{Coefficients for perturbative correction, $\delta_P^{(1)}(s,l)$
 for each pair of $(s,l)$.
 The square brackets in the first column indicate the fit range in $u$.
 }
 \label{tab:deltaP}
\end{table}
%%%%%%%%%%%%%%%%%%%%%%%

\begin{figure}[tb]
\centering
\begin{tabular}{cc}
\includegraphics*[width=0.5 \textwidth,clip=true]
{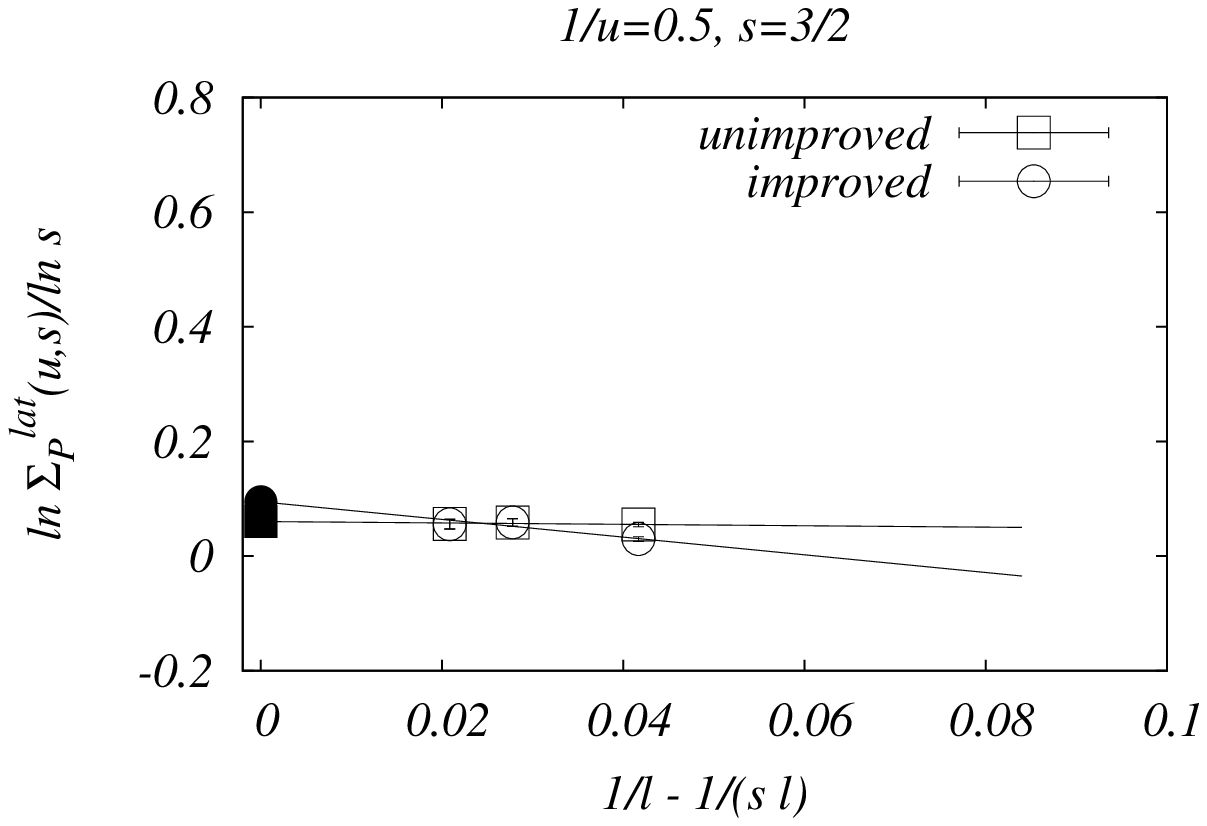}&
\includegraphics*[width=0.5 \textwidth,clip=true]
{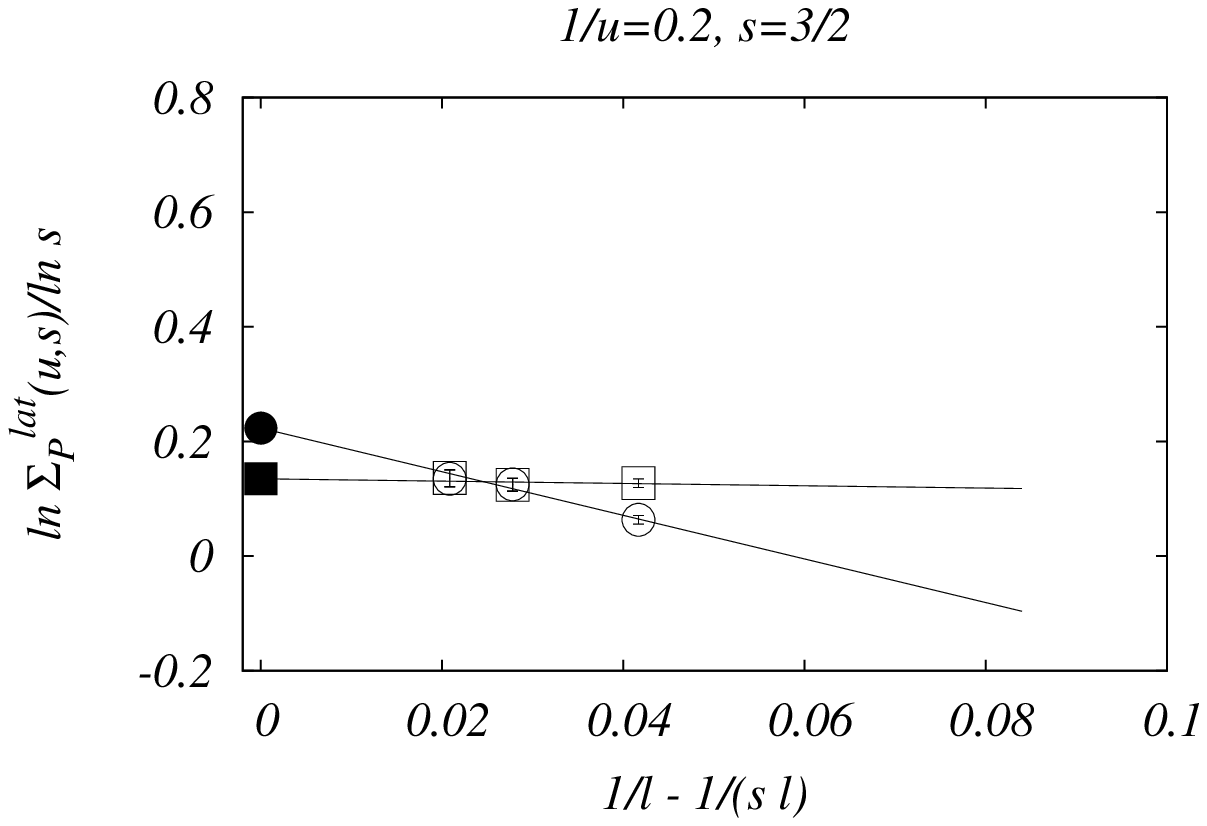}\\
\includegraphics*[width=0.5 \textwidth,clip=true]
{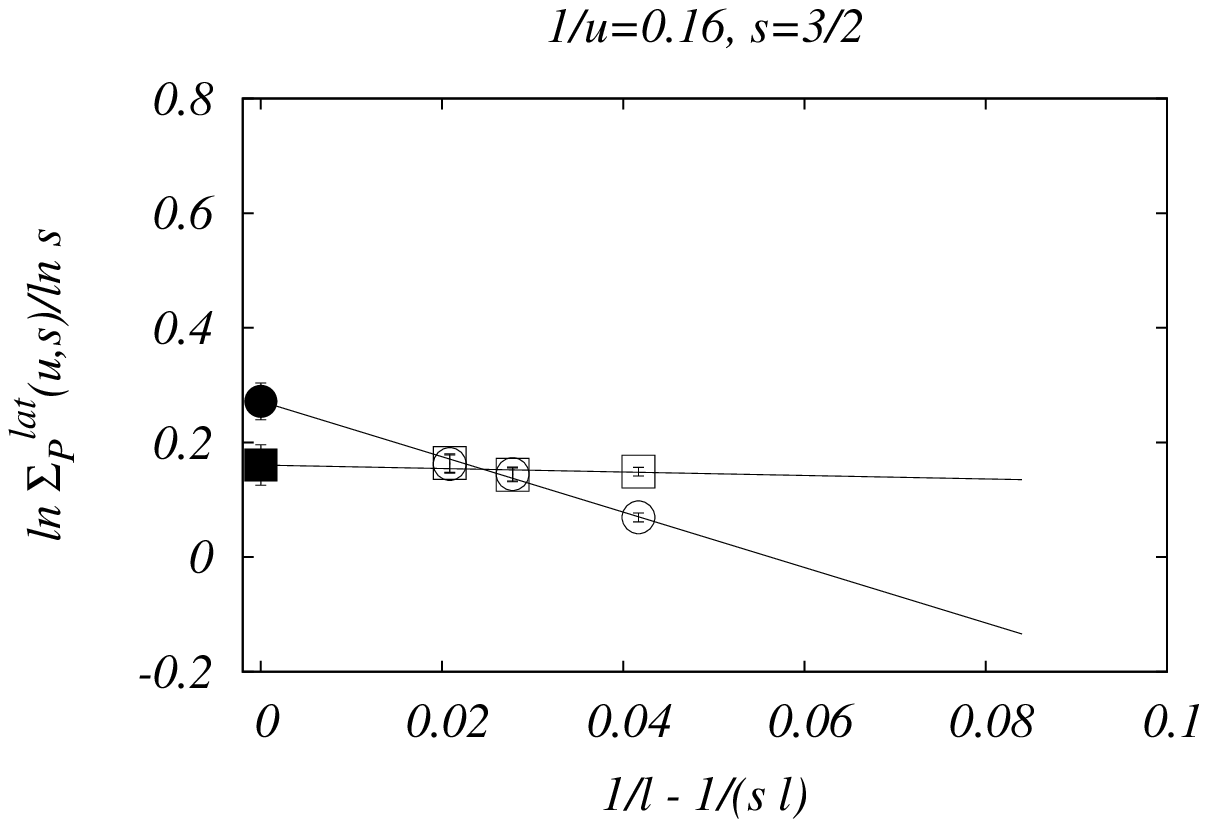}&
\includegraphics*[width=0.5 \textwidth,clip=true]
{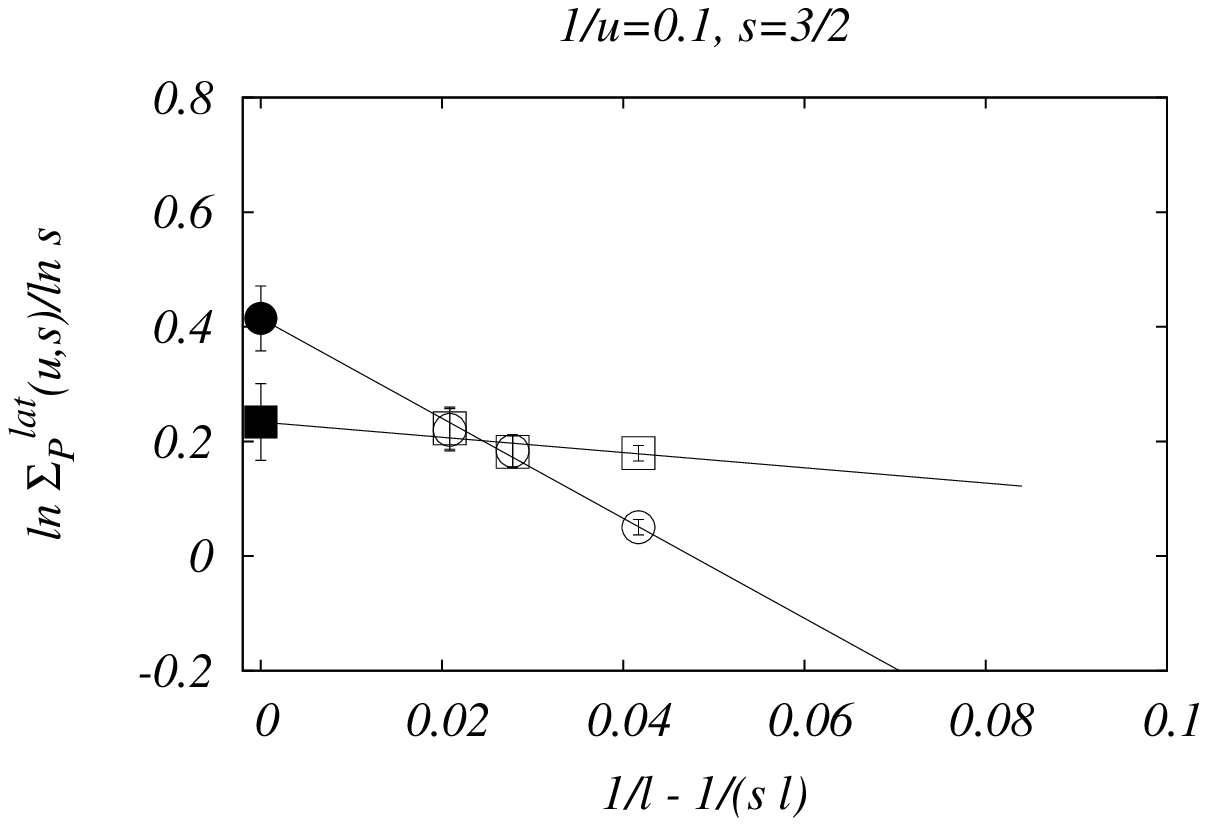}\\
\end{tabular}
\caption{Continuum limit of $\ln \sigma_P(u,s)/\ln s$ for $s=1.5$.}
\label{fig:gamma_m_contlim-s1.5}
\end{figure}

\begin{figure}[tb]
\centering
\begin{tabular}{cc}
\includegraphics*[width=0.5 \textwidth,clip=true]
{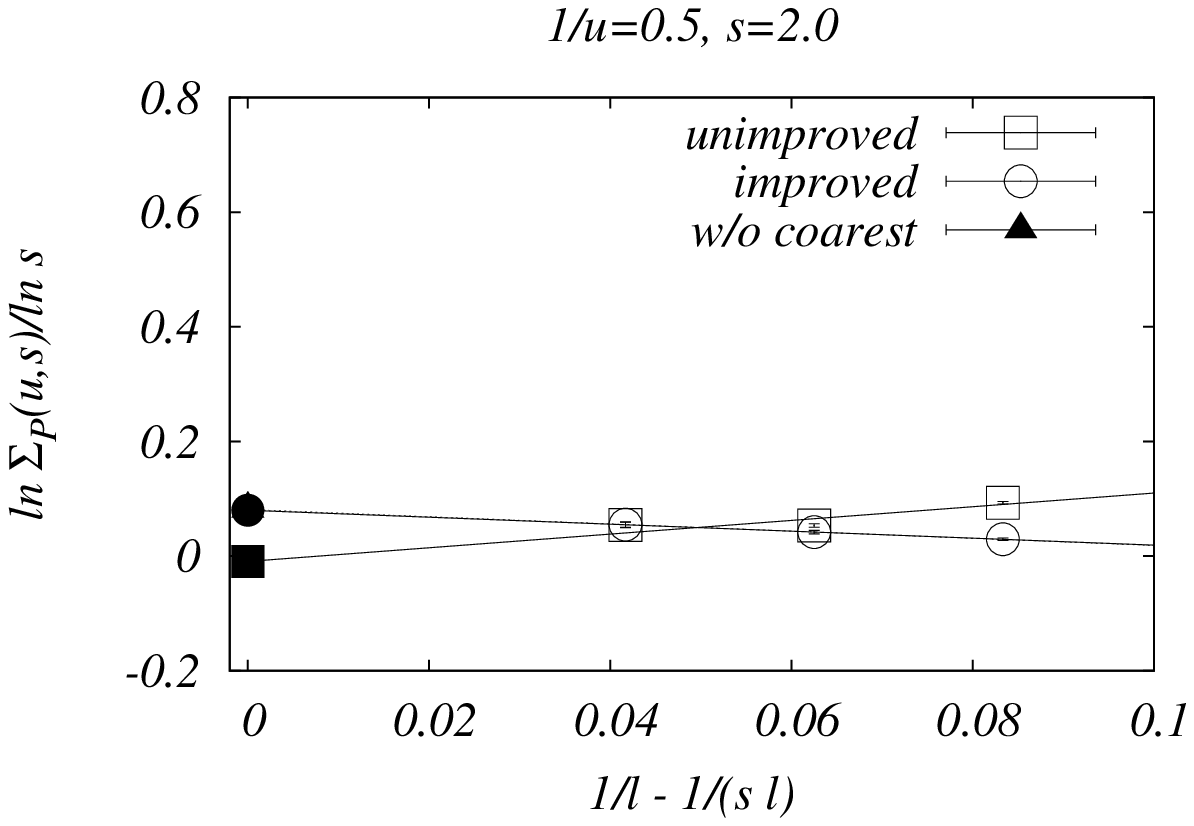}&
\includegraphics*[width=0.5 \textwidth,clip=true]
{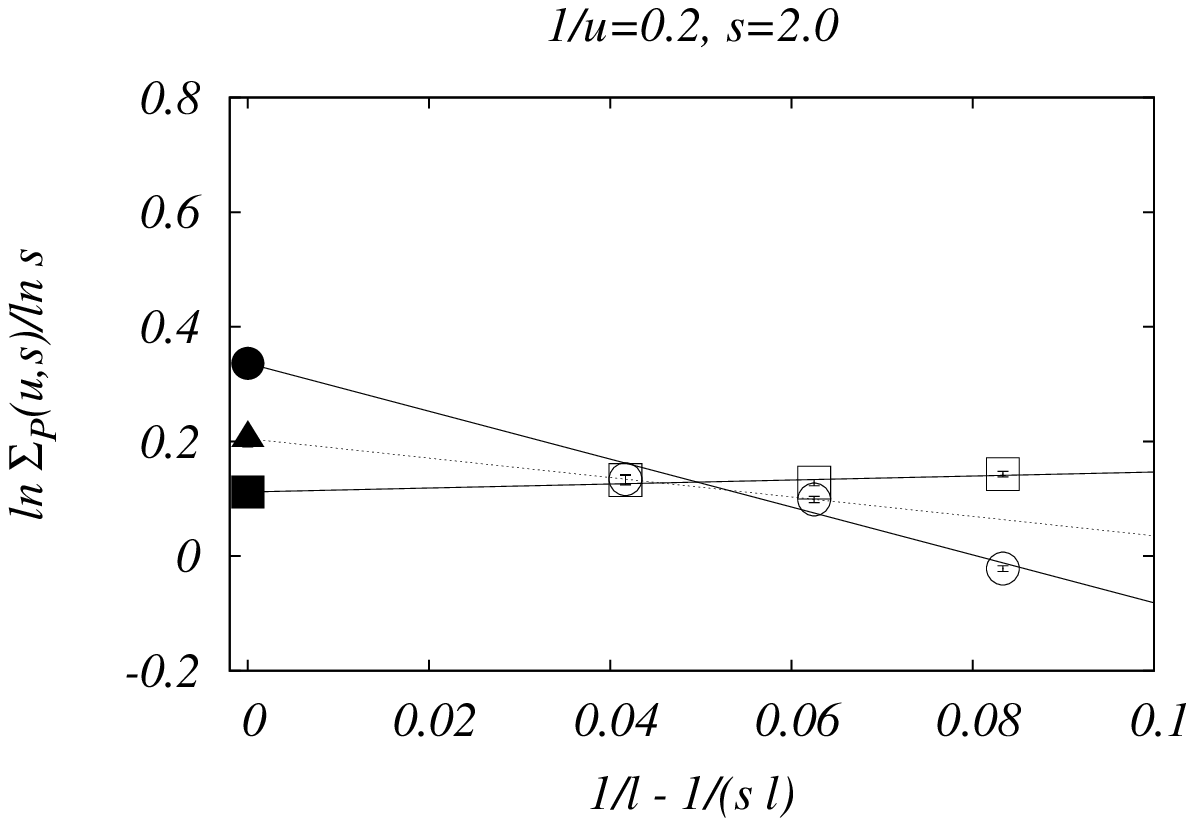}\\
\includegraphics*[width=0.5 \textwidth,clip=true]
{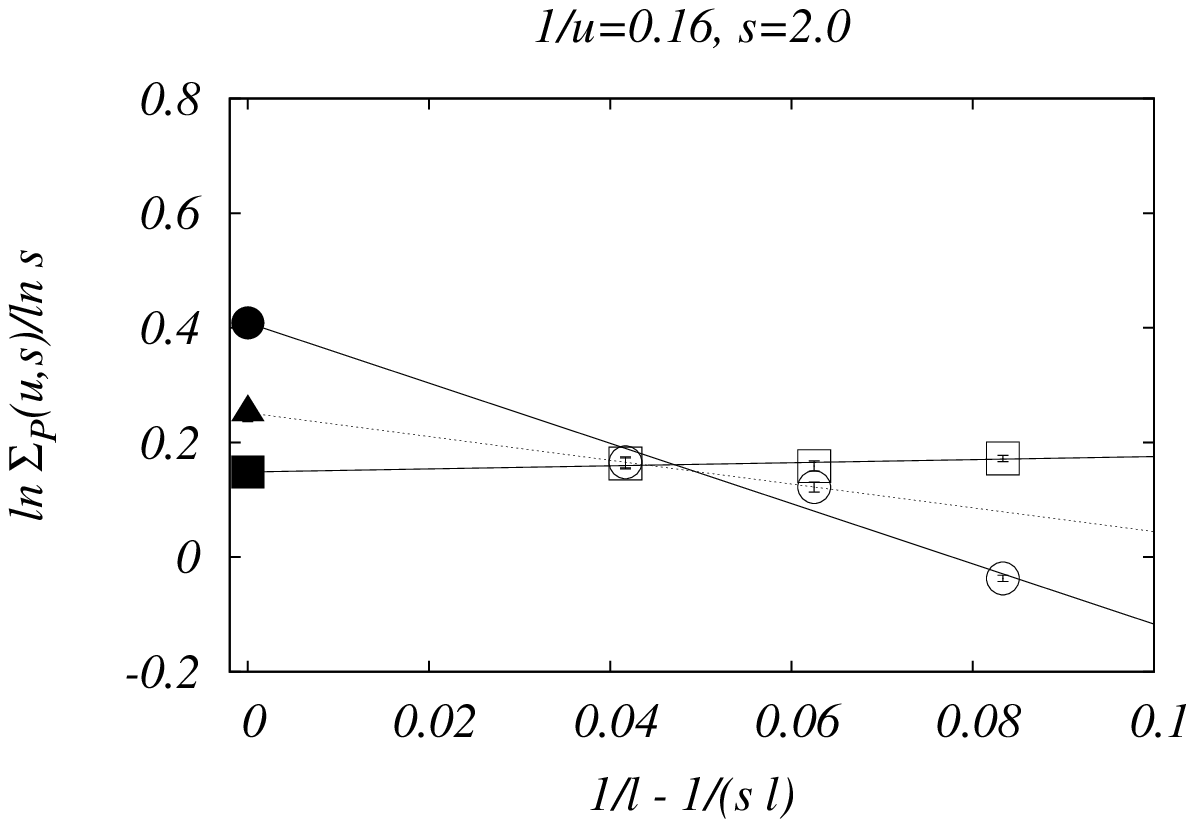}&
\includegraphics*[width=0.5 \textwidth,clip=true]
{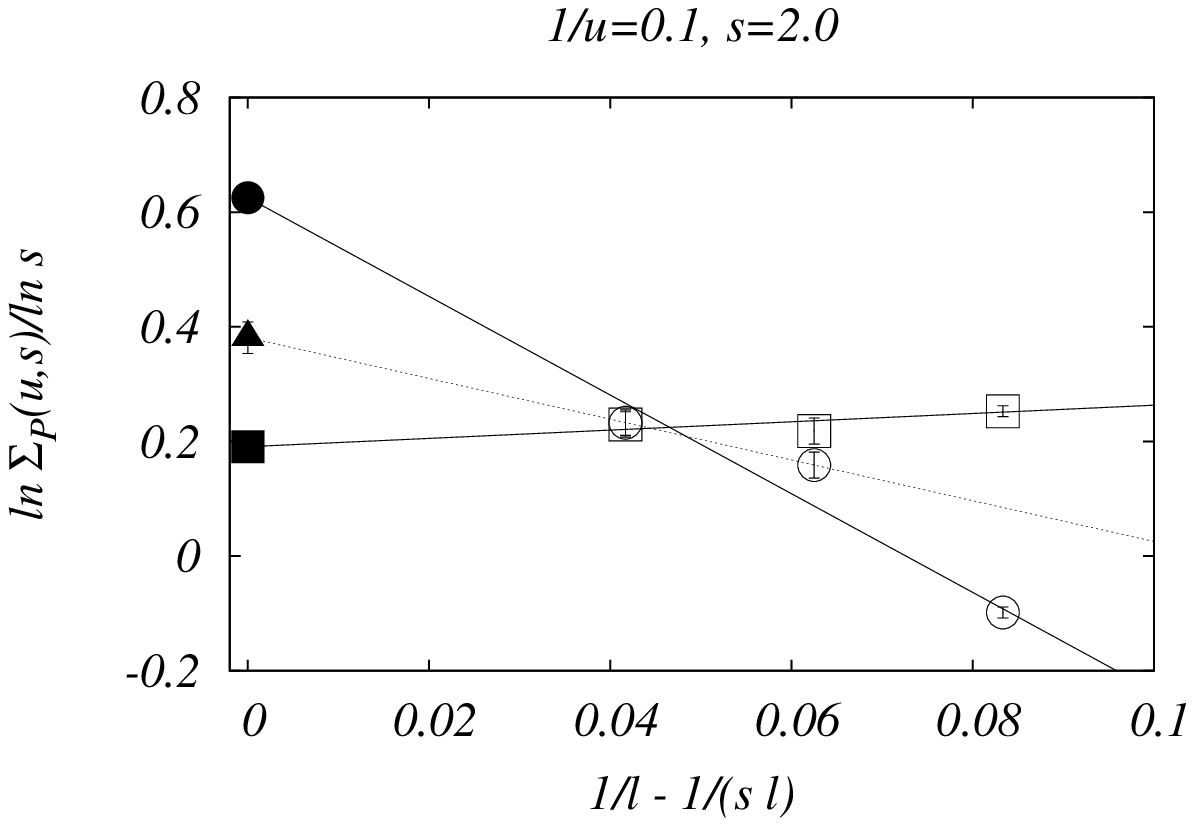}\\
\end{tabular}
\caption{Continuum limit of $\ln \sigma_P(u,s)/\ln s$ for $s=2$.}
\label{fig:gamma_m_contlim-s2.0}
\end{figure}

Once the improvement coefficient is determined, the improved step
scaling function for $Z_P$ is obtained by
\begin{eqnarray}
  \Sigma_{P,1}^{\rm lat}(u,s,l)
= \frac{\Sigma_{P,0}^{\rm lat}(u,s,l)}{1 + \delta_P^{(1)}(s,l) u}.
\end{eqnarray}
The continuum limit of $\Sigma_{P,i}^{\rm lat}$ ($i=0$ or 1) is taken
with a fixed $s$(=3/2 or 2) for the various initial value of the
coupling constant $u$.
In practice, the continuum limit is taken for $\ln \sigma_P(u,s)/\ln s$
introduced in Eq.~(\ref{eq:gamma_m_star}) with a linear function in $a$.
Figures \ref{fig:gamma_m_contlim-s1.5} and \ref{fig:gamma_m_contlim-s2.0}
show the continuum limit of $\ln \sigma_P(u,s)/\ln s$ with and without
the improvement at the four representative values of $1/u$ at $s=3/2$
and 2, respectively.
%%%%%%%%%%%%%%%%%%%%%%%
%%%%%%%%%%%%%%%%%%%%%%%
%%%%%%%%%%%%%%%%%%%%%%%
%%%%%%%%%%%%%%%%%%%%%%%

\begin{figure}[tb]
\centering
\begin{tabular}{c}
\includegraphics*[width=0.7 \textwidth,clip=true]
{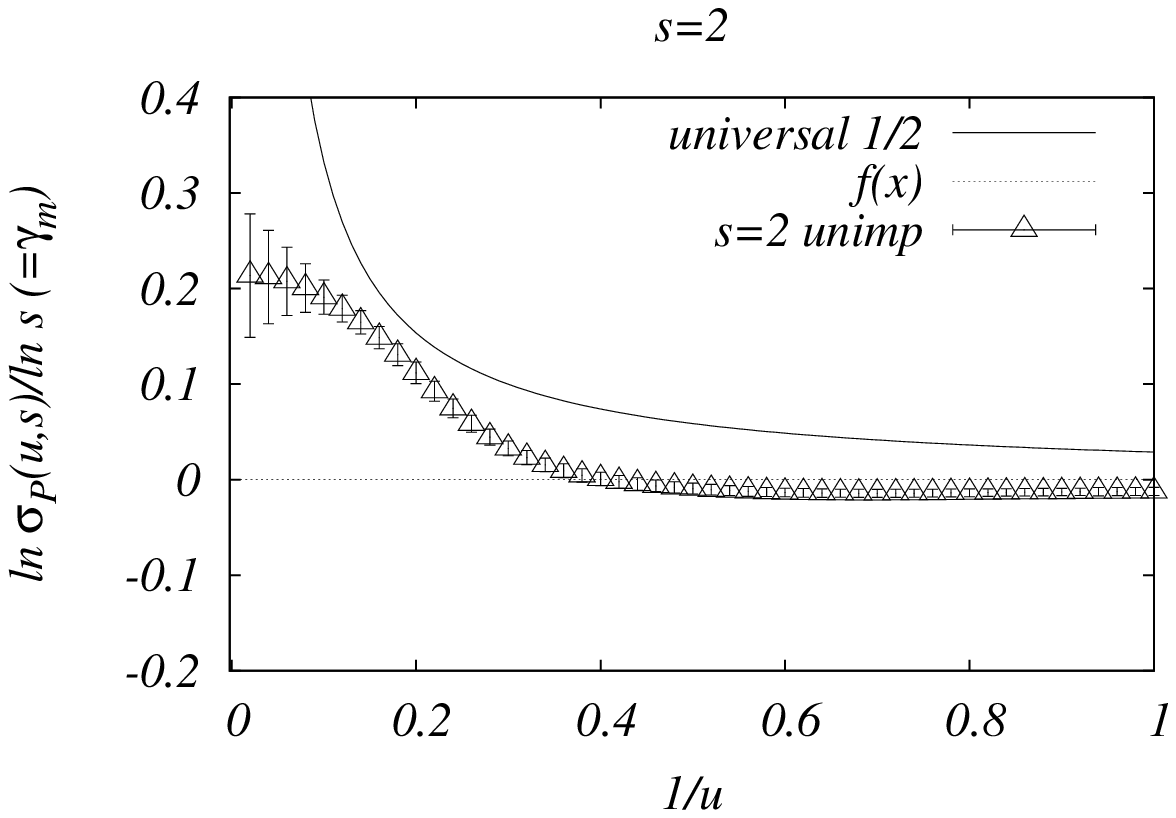}\\
\end{tabular}
\caption{Continuum limit of $\gamma_m$ at $s=2$ without improvement.}
\label{fig:gamma_m-imp-unimp}
\end{figure}

Looking at these figures carefully, one question might arise.
Namely, the unimproved data appears to have a scaling violation smaller
than the improved one.
In Fig.~\ref{fig:gamma_m-imp-unimp}, we plot the $1/u$ dependence of the
continuum limit without improvement for $s=2$.
As the figure clearly shows, in the small coupling region, where
perturbation is reliable, the continuum limit is much smaller than the
perturbative prediction.
From this observation, we infer that the small scaling violation for the
unimproved data is fake and that the improvement removes such nonlinear
discretization errors efficiently and makes the data align.
In the following, we only analyze the improved data.
%%%%%%%%%%%%%%%%%%%%%%%
%%%%%%%%%%%%%%%%%%%%%%%

In Fig.~\ref{fig:gamma_m_contlim-s1.5}, the three improved data points
at $s=3/2$ well align independently of the value of $1/u$, while it is
hard to justify the linear extrapolation using three data for $s=2$ as
seen in Fig.~\ref{fig:gamma_m_contlim-s2.0}.
Thus, at $s=2$ we omit the coarsest data point and take the continuum
limit as before.
The fit results without the coarsest point are shown in
Fig.~\ref{fig:gamma_m_contlim-s2.0}.

\begin{figure}[tb]
\centering
\begin{tabular}{c}
\includegraphics*[width=0.7 \textwidth,clip=true]
{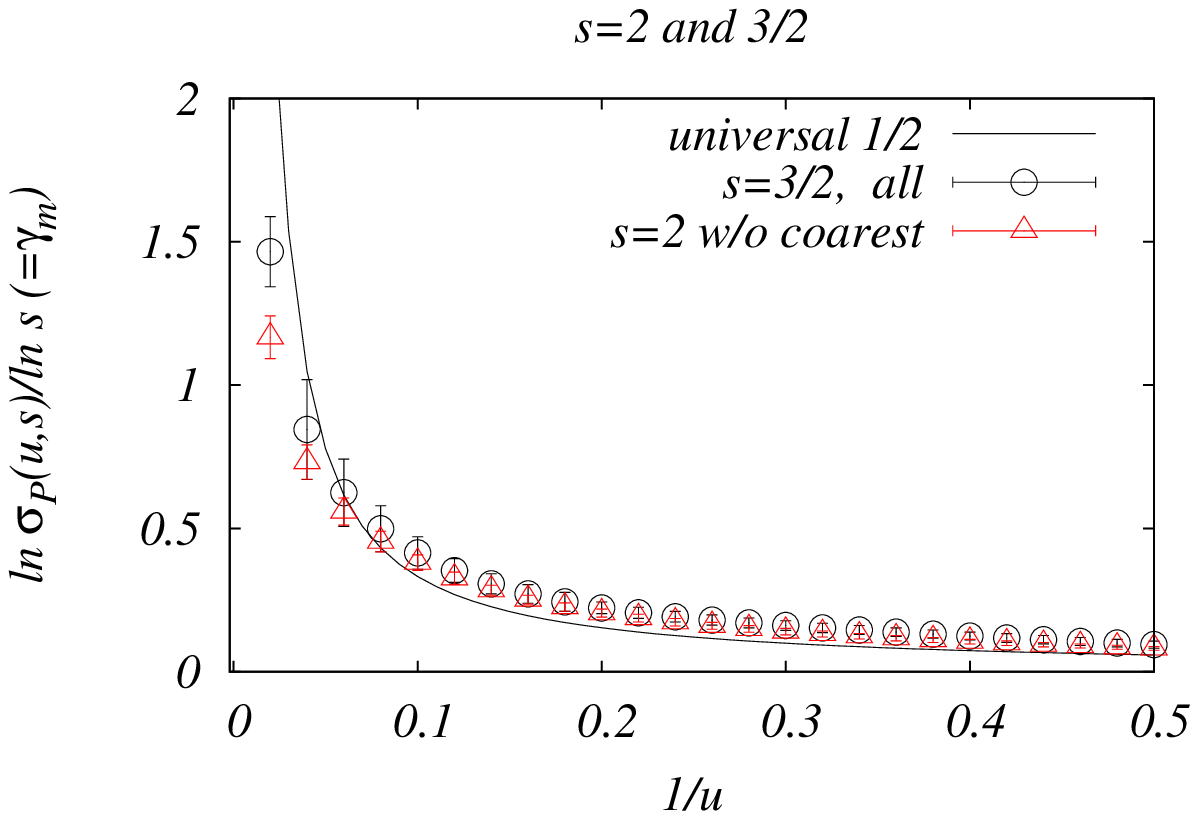}\\
\end{tabular}
\caption{Continuum limit of $\gamma_m$.}
\label{fig:gamma_m}
\end{figure}

$\ln \sigma_P(u,s)/\ln s$ depends on $s$, but if the IRFP exists its
value becomes $\gamma_m^*$ independently of $s$ at the IRFP.
To check this, we plot $\ln \sigma_P(u,s)/\ln s$ for $s$=3/2 and 2 as a
function of $1/u$ in Fig.~\ref{fig:gamma_m}.
The perturbative prediction shown in the figure is calculated using the
combination of the two-loop $\beta$ function and the one-loop
anomalous dimension for $s=2$, and that for $s=3/2$ is omitted because
the difference is too tiny to distinguish.

%%%%%%%%%%%%%%%%%%%%%%%
%%%%%%%%%%%%%%%%%%%%%%%
Recalling the range of the possible IRFP value $0.06\le 1/u \le 0.15$,
it is seen that two results agree with each other in that range, which
is interpreted as another support to the existence of the IRFP and
justify our analysis.
Provided that the value of IRFP is in $0.06 \lesssim 1/u \lesssim 0.15$,
it is found that $0.26 \lesssim \gamma^*_m \lesssim 0.74$.
%$0.29(3) \lessim \gamma_m \lessim 0.62(12)$

\section{Summary and discussion}
\label{sec:discussion}

In this work, the running coupling constant and the mass anomalous
dimension of six-flavor two-color QCD are numerically investigated using
the lattice step scaling technique.
The discretization errors are improved perturbatively and the
improvement turns out to decrease the DBF, which explains the
discrepancy between Ref.~\cite{Bursa:2010xn} and
Ref.~\cite{Karavirta:2011zg} qualitatively.
The extrapolation of the DBF to the continuum limit is taken linearly
assuming that the $O(a)$ scaling violation dominates the higher order
ones.
The DBF in the continuum limit turns out to approach zero from below as
the inverse SF coupling constant $1/u$ decreases, and it becomes
consistent with zero in the range $0.06 \le 1/u \le 0.15$.
The linear extrapolation is reasonably justified within the statistical
error, but a further rigorous check is clearly preferable.
The result of this work suggests that SU(2) gauge theory with six Dirac
fermions in the fundamental representation is in the conformal window.

One possible loophole in our analysis may be related to the validity of
the linear extrapolation.
Since the number of data points used in the continuum extrapolation is
rather limited, we cannot check the validity rigorously.
Within this limitation, we have made a nontrivial check that the
continuum limits of the DBF or $\ln\sigma_P/\ln s$ with two different
values of $s$ agree.
Such a check sustains the possibility that the systematic uncertainty
due to neglecting $O(a^2)$ or higher order scaling violations is not
significantly large.

In order to confirm the existence of the IRFP or even determine the more
precise value of the fixed point, data from larger lattices with high
statistics are necessary.
It is, however,  difficult to do with machines currently available to
us, and probably more efficient methods or approaches are necessary to
go further.
The conformal window can also be studied by looking at hadron
spectroscopy or renormalization group analysis on the lattice.
In order to establish the location of the lower end of the conformal
window, the consistency check employing these methods would be
indispensable.
What is the most important in the context of the WTC is the value of the
anomalous dimension of the $\overline{\psi} \psi$ operator, and it is
found that $0.26 \lesssim \gamma^*_m \lesssim 0.74$ in the possible
range of the IRFP.

The anomalous dimension of the six-flavor theory turns out to be smaller
than the value required in phenomenology, and hence the five- or
four-flavor theory may be interesting to study.
Provided that one eventually succeeds to find an attractive candidate
for WTC through conformal window search, the next step would be the
calculation of the $S$ parameter.
The calculational method has been established in
Ref.~\cite{Shintani:2008qe}, where the QCD $S$ parameter is calculated
on the lattice for the first time and is correctly reproduced.
In Ref.~\cite{Appelquist:2010xv}, the evidence of the reduction of
$S$ parameter is observed in the presence of many flavors.
Another important direction is obviously to calculate the mass spectrum
of the candidate theories, including vector and scalar resonances, the
decay constant of the Nambu Goldstone boson, and the chiral condensate.
Our first result is reported in Ref.~\cite{Hayakawa:2013maa}.
Although the precise determinations of these quantities are challenging,
the direct comparison with the upcoming LHC results is extremely
interesting, and we believe that such calculations are worth a lot of
effort.

\section{Acknowledgments}

N.Y. would like to thank Hideo Matsufuru for useful discussions as well
as providing us with a comfortable machine environment.
The main part of the numerical simulations were performed on the
B-factory computer system at KEK and on the computer system $\varphi$ at
Nagoya University.
This work is supported in part by the Grant-in-Aid for Scientific
Research of the Japanese Ministry of Education, Culture, Sports,
Science, and Technology and JSPS
(No. 20540261, % hayakawa
 No. 22224003, % hayakawa
 No. 22740183, % yamada wakate B
 and
 No. 23740177  % takeda
).

\appendix
\section{Perturbative coefficients}

\label{sec:four-loop-coefficients}

\subsection{$\beta$ function}

In the MS (and $\msbar$ scheme), the coefficients of the $\beta$
function for SU($N_c$) gauge theory with $N_f$ flavors in the
fundamental representation are known to the four-loop
level~\cite{vanRitbergen:1997va} as
\begin{eqnarray}
     b_1
 &=& \frac{2}{\left(4\pi\right)^2}
     \left[ \frac{11 C_A}{3} - \frac{4 T_F}{3}N_f
     \right],\\
     b_2
 &=& \frac{2}{\left(4\pi\right)^4}
     \left[\, \frac{34 C_A^2}{3}
            - \left( \frac{20 C_A}{3} + 4 C_F \right) T_F N_f\,
     \right],\\
     b_3^{\msbar}
 &=& \frac{2}{\left(4 \pi\right)^6}
     \Bigg[   \frac{2857}{54} C_A^3
           + \left(   2 C_F^2 - \frac{205}{9} C_F C_A
                    - \frac{1415}{27} C_A^2         \right) T_F N_f
           + \left(  \frac{44}{9} C_F + \frac{158}{27} C_A \right)
             T_F^2 N_f^2
     \Bigg],\no\\ \\
     b_4^{\msbar}
 &=& \frac{2}{\left(4 \pi\right)^8}
     \Bigg[   C_A^4 \left( \frac{150653}{486} - \frac{44}{9}\, \zeta_3
                    \right)
           + C_A^3       T_F   N_f
             \left( - \frac{39143}{81} + \frac{136}{3}\, \zeta_3
             \right)
\no\\&&\hspace*{8ex}
           + C_A^2 C_F   T_F   N_f
             \left(\frac{7073}{243} - \frac{656}{9}\, \zeta_3
             \right)
           + C_A   C_F^2 T_F   N_f
             \left( - \frac{4204}{27} + \frac{352}{9}\, \zeta_3
             \right)
\no\\&&\hspace*{8ex}
           + 46    C_F^3 T_F N_f
           + C_A^2       T_F^2 N_f^2
             \left(\frac{7930}{81} + \frac{224}{9}\, \zeta_3
             \right)
           +       C_F^2 T_F^2 N_f^2
             \left(\frac{1352}{27} - \frac{704}{9}\, \zeta_3
             \right)
\no\\&&\hspace*{8ex}
           + C_A   C_F   T_F^2 N_f^2
             \left(\frac{17152}{243} + \frac{448}{9}\, \zeta_3
             \right)
           + \frac{424}{243} C_A T_F^3 N_f^3
           + \frac{1232}{243} C_F T_F^3 N_f^3
\no\\&&\hspace*{8ex}
           + N_c^2 \frac{N_c^2 + 36}{24}
             \left(-\frac{80}{9} + \frac{704}{3}\, \zeta_3
             \right)
           + N_f N_c \frac{N_c^2 + 6}{48}
             \left(\frac{512}{9} - \frac{1664}{3}\, \zeta_3
             \right)
\no\\&&\hspace*{8ex}
           + N_f^2 \frac{N_c^4 - 6 N_c^2 + 18}{96 N_c^2}
             \left(- \frac{704}{9} + \frac{512}{3}\, \zeta_3
             \right)
     \Bigg],
\end{eqnarray}
where
\begin{eqnarray}
\zeta_3=   1.202056903,\ \
\zeta_4=   1.0823232,\ \
\zeta_5=   1.0369277,
\end{eqnarray}
and $T_F = \frac{1}{2}$.
For $N_c=2$,
\begin{eqnarray}
&& C_A = N_c = 2,\ \
   C_F = \frac{N_c^2 - 1}{2 N_c} = \frac{3}{4}.
% N_A=N_c^2 - 1,
%T_A=N_c,
%d_F=\frac{\left(N_c^4 - 6 N_c^2 + 18}{48 N_c^2},
%d_A=N_c \frac{N_c^2 + 6}{24},
\end{eqnarray}

\subsection{Anomalous dimension}

In the MS (and $\msbar$ scheme), the coefficients of the mass anomalous
dimension for SU($N_c$) gauge theory with $N_f$ flavors in the
fundamental representation are known to the four-loop
level~\cite{Vermaseren:1997fq} as
\begin{eqnarray}
     d_1
 &=& \frac{2}{\left(4\pi\right)^2}\ 3\, C_F,\\
     d_2
 &=& \frac{2}{\left(4 \pi\right)^4}
     \Big[   \frac{3}{2} C_F^2 + \frac{97}{6} C_F\,C_A
           - \frac{10}{3}C_F\,T_F\,N_f
     \Big],
\\
     d_3
 &=& \frac{2}{\left(4 \pi\right)^6}
     \Bigg[   \frac{129}{2} C_F^3 - \frac{129}{4} C_F^2 C_A
           + \frac{11413}{108} C_F\,C_A^2
           + C_F^2\, T_F\, N_f \left( - 46 + 48\, \zeta_3 \right)
\no\\&&\hspace*{8ex}
           + C_F\, C_A\, T_F\, N_f \left(-\frac{556}{27} - 48\, \zeta_3
                                   \right)
           - \frac{140}{27} C_F\, T_F^2\, N_f^2
     \Bigg],\\
     d_4
 &=& \frac{2}{\left(4 \pi\right)^8}
     \Bigg[   C_F^4 \left( - \frac{1261}{8} - 336\, \zeta_3 \right)
           + C_F^3 C_A \left(\frac{15349}{12} + 316\, \zeta_3 \right)
\no\\&&\hspace*{8ex}
           + C_F^2 C_A^2 \left(- \frac{34045}{36} - 152\, \zeta_3
                               + 440\, \zeta_5
                         \right)
           + C_F C_A^3
             \left(   \frac{70055}{72} + \frac{1418}{9}\, \zeta_3
                    - 440\, \zeta_5 \right)
\no\\&&\hspace*{8ex}
           + C_F^3 T_F N_f
             \left(-\frac{280}{3} + 552\, \zeta_3 - 480\, \zeta_5 \right)
\no\\&&\hspace*{8ex}
           + C_F^2 C_A T_F N_f
             \left( - \frac{8819}{27} + 368\, \zeta_3
               - 264\, \zeta_4 + 80\, \zeta_5 \right)
\no\\&&\hspace*{8ex}
           + C_F C_A^2 T_F N_f
             \left( - \frac{65459}{62} - \frac{2684}{3}\, \zeta_3
               + 264\, \zeta_4 + 400\, \zeta_5 \right)
\no\\&&\hspace*{8ex}
           + C_F^2 T_F^2 N_f^2
             \left(\frac{304}{27} - 160\, \zeta_3 + 96\, \zeta_4
             \right)
           + C_F C_A T_F^2 N_f^2
             \left(\frac{1342}{81} + 160\, \zeta_3 - 96\, \zeta_4
             \right)
\no\\&&\hspace*{8ex}
           + C_F T_F^3 N_f^3
             \left(-\frac{664}{81} + \frac{128}{9}\, \zeta_3 \right)
           + \frac{\left(N_c^2 -1\right)\left(N_c^2 + 6\right)}{48}
             \left(-32 + 240\, \zeta_3\right)
\no\\&&\hspace*{8ex}
           + N_f \frac{\left(N_c^2-1\right)
             \left(N_c^4 - 6 N_c^2 + 18\right)}{96 N_c^3}
             \left( 64 - 480\, \zeta_3 \right)
     \Bigg].
\end{eqnarray}

\section{\SF scheme}
\label{sec:sf-setup}

The SF on the lattice is defined on a four-dimensional hypercubic
lattice with a volume $(L/a)^4$ in the cylindrical geometry, and the
physical length $L$ is identified as the renormalization scale.
The periodic boundary condition in the spatial directions with a
vanishing phase factor ($\theta=0$) and the Dirichlet one in the
temporal direction are imposed for both gauge [$U_\mu(x)$] and fermion
[$\psi(x)$ and $\bar\psi(x)$] fields.
The boundary values for gauge and fermion fields are represented by
two-by-two color matrices, $C$ and $C'$, and spinors, $\rho$,
$\rho'$, $\bar\rho$, and $\bar\rho'$, respectively.
The partition function of this system is given by
\begin{eqnarray}
 Z_{\rm SF}(C',\bar \rho',\rho'\,; C,\bar \rho,\rho)
=e^{-\Gamma(C',\bar \rho',\rho'\,; C,\bar \rho,\rho)}
=\int D[U,\psi,\bar\psi]
 e^{-S[U,\psi,\bar\psi,C,C',\rho,\rho',\bar \rho,\bar \rho']},
\end{eqnarray}
where $\Gamma$ is the effective action, and
\begin{equation}
S[U,\psi,\bar\psi,C,C',\rho,\rho',\bar \rho,\bar\rho']
=S_g[U,C,C']+S_q[U,\psi,\bar\psi,\rho,\rho',\bar\rho,\bar\rho']. 
\end{equation}
We take the plaquette gauge action,
\begin{eqnarray}
S_g[U,C,C'] = \frac{\beta}{4}
         \sum_{x}\sum_{\mu=0}^3\sum_{\nu=0}^3
         \bar\delta_{\mu,\nu}w_{\mu,\nu}(x_0)~
         {\rm Tr}\left[1-P_{\mu,\nu}(x) \right],
\label{eq:gauge action}
\end{eqnarray}
where $\beta=4/g_0^2$ set the lattice bare coupling constant $g_0^2$,
$\bar\delta_{\mu,\nu}$=0 when $\mu=\nu$ otherwise 1, and
$P_{\mu,\nu}(x)$ denotes a 1$\times$1 Wilson loop on the $\mu$-$\nu$
plane starting and ending at $x$.
The spatial link variables at $x_0=0$ and $L/a$ are all set to the
diagonal, constant matrices~\cite{Luscher:1992zx},
\begin{eqnarray}
  U_k(x)|_{x_0=0}
= \exp \left[ C \right],&&
  C
= \frac{a}{i L} \left(
  \begin{array}{cc}
  \eta &     0 \\
     0 & -\eta \\
  \end{array} \right),
\label{formul:BGF:C0}\\
  \left. U_k(x)\right|_{x_0=L/a}
= \exp \left[ C' \right],&&
  C'
= \frac{a}{i L} \left(
  \begin{array}{cc}
   \pi-\eta &     0     \\
      0     & -\pi+\eta \\
  \end{array} \right)
\label{formul:BGF:CT},
\end{eqnarray}
where $k=1$, 2, 3, and $\eta$ is parametrizing the gauge boundary
fields and is eventually set to $\pi/4$.
The weight $w_{\mu,\nu}(x_0)$ in Eq.~(\ref{eq:gauge action}) is given by
\begin{eqnarray}
w_{\mu,\nu}(x_0) &=& \left\{\begin{array}{cl}
  c_t            & \mbox{for ($t=0$ or $t=(L/a)-1$) and
                         ($\mu$ or $\nu$=0)}\\ 
  0              & \mbox{for ($t=(L/a)$) and ($\mu$ or $\nu$=0)}\\
  \frac{1}{2}c_s & \mbox{for ($t=0$ or $t=(L/a)$) and
                         ($\mu\ne$0 and $\nu\ne$0)}\\
  1              & \mbox{for all the other cases}
                  \end{array}\right..
\label{eq:wmunu}
\end{eqnarray}
By tuning $c_t$, $O(a)$ errors induced by the boundaries in the time
direction can be removed perturbatively, but in this work we simply take
its tree level values, $c_t=1$.
With this setup, the value of $c_s$ is arbitrary because the spatial
plaquettes on the boundaries do not contribute to the action.
We thus set $c_s=0$.

The fermion fields are described by the unimproved Wilson fermion action,
\begin{eqnarray}
     S_q[U,\psi,\bar\psi]
 &=& N_f \sum_{x,y} \bar{\psi}(x) D(x,y;U) \psi(y)\,
  =  N_f \sum_{x,y} \bar{\psi}^{\rm lat}(x) D^{\rm lat}(x,y;U) \psi^{\rm lat}(y),
   \label{eqn:setup:Sq1}\\
     D^{\rm lat}(x,y;U)
 &=& \delta_{xy} 
    -\kappa \sum_{\mu} 
     \left\{ \left( 1 - \gamma_{\mu} \right)
             U_\mu(x) \delta_{x+\hat{\mu},y}
           + \left( 1 + \gamma_{\mu} \right)
             U^{\dagger} _\mu(x-\hat \mu)
             \delta_{x-\hat{\mu},y}
     \right\},
   \label{eqn:setup:Sq2}
\end{eqnarray}
where
\begin{eqnarray}
    \psi^{\rm lat}(x)
&=& \frac{1}{\sqrt{2\kappa}}\, \psi(x),\ \ \
    \bar \psi^{\rm lat}(x)
 =  \frac{1}{\sqrt{2\kappa}}\, \bar \psi(x),\ \ \
    D^{\rm lat}(x,y;U)
 =  2\kappa\, D(x,y;U)\,.
\end{eqnarray}
The hopping parameter $\kappa$ is related to the dimensionless bare mass
$M_0$ through $2\,\kappa=1/(M_0 + 4)$.
The dynamical degrees of freedom of the fermion field $\psi(x)$ and
antifermion fields $\bar\psi(x)$ reside on the lattice sites $x$ with
$0<x_0<T$.
On both boundaries ($x_0=0$ and $T$), half of the Dirac components
are set to zero and the remaining components are fixed to some
prescribed values, $\rho$, $\bar\rho$, $\rho'$, and $\bar\rho'$, as
\begin{eqnarray}
 \left.P_+\psi(x)\right|_{x_0=0}=\rho({\bf x}),\ \ \
 \left.P_-\psi(x)\right|_{x_0=0}=0,\\
 \left.P_-\psi(x)\right|_{x_0=T}=\rho'({\bf x}),\ \ \
 \left.P_+\psi(x)\right|_{x_0=T}=0,\\
 \left.\bar\psi(x)P_-\right|_{x_0=0}=\bar\rho({\bf x}),\ \ \
 \left.\bar\psi(x)P_+\right|_{x_0=0}=0,\\
 \left.\bar\psi(x)P_+\right|_{x_0=T}=\bar\rho'({\bf x}),\ \ \
 \left.\bar\psi(x)P_-\right|_{x_0=T}=0,
\end{eqnarray}
where $P_\pm=(1\pm\gamma_0)/2$.
In this work, the boundary values for the fermion fields are set to
zero, {\it i.e.},
\begin{eqnarray}
\rho=\rho'=\bar\rho=\bar\rho'=0.
\end{eqnarray}

\section{Raw data}
\label{sec:raw-data}

\begin{table}
 \begin{tabular}{rccrcccr}
$\beta$ & $\kappa$ & Trajs. & plq. & $\delta\tau$ & Acc. & $g_{\rm SF}^2$ & $M$ \\
\hline
10.00 & 0.1295040 & 30,000 & 0.926136(16) &0.1429 &0.789(2) & 0.44518(41) &$ 0.00007$(4)\\
\hline
9.00 & 0.1299100 & 30,000 & 0.917889(12) &0.1429 &0.824(2) & 0.50087(63) &$-0.00005$(4)\\
\hline
8.00 & 0.1304270 & 30,000 & 0.907552(12) &0.1667 &0.748(3) & 0.57331(81) &$ 0.00007$(9)\\
\hline
7.50 & 0.1307300 & 30,000 & 0.901344(14) &0.1667 &0.773(2) & 0.61855(86) &$ 0.00045$(10)\\
\hline
7.00 & 0.1311050 & 30,000 & 0.894285(18) &0.1667 &0.788(3) & 0.66875(70) &$ 0.00002$(6)\\
\hline
6.00 & 0.1320476 & 30,000 & 0.876503(19) &0.1667 &0.815(2) & 0.8055(11) &$ 0.00015$(7)\\
\hline
5.50 & 0.1326700 & 30,000 & 0.865107(37) &0.1667 &0.824(3) & 0.8962(17) &$ 0.00012$(10)\\
\hline
5.00 & 0.1334600 & 30,000 & 0.851424(25) &0.1667 &0.840(3) & 1.0101(20) &$-0.00012$(13)\\
\hline
4.50 & 0.1344250 & 45,000 & 0.834738(23) &0.2000 &0.750(2) & 1.1605(26) &$ 0.00013$(13)\\
\hline
4.00 & 0.1357000 & 30,000 & 0.813595(32) &0.2000 &0.774(3) & 1.3564(24) &$ 0.00048$(21)\\
\hline
3.50 & 0.1375000 & 30,000 & 0.786531(51) &0.2000 &0.794(2) & 1.6297(69) &$-0.00060$(23)\\
\hline
3.00 & 0.1400480 & 60,000 & 0.750086(30) &0.2000 &0.814(2) & 2.0760(76) &$ 0.00078$(15)\\
\hline
2.50 & 0.1441900 & 60,000 & 0.699079(34) &0.2000 &0.820(2) & 2.829(12) &$-0.00022$(24)\\
\hline
2.40 & 0.1453500 & 60,000 & 0.686330(35) &0.2000 &0.820(2) & 3.095(13) &$-0.00050$(26)\\
\hline
2.30 & 0.1466400 & 60,000 & 0.672451(28) &0.2000 &0.824(1) & 3.356(15) &$-0.00045$(25)\\
\hline
2.20 & 0.1481000 & 60,000 & 0.657554(42) &0.2000 &0.822(2) & 3.776(24) &$ 0.00033$(30)\\
\hline
2.10 & 0.1498190 & 60,500 & 0.641272(58) &0.2000 &0.825(2) & 4.222(31) &$-0.00029$(38)\\
\hline
2.00 & 0.1517700 & 60,000 & 0.623513(50) &0.2000 &0.823(2) & 4.867(39) &$ 0.00030$(63)\\
\hline
1.90 & 0.1541500 & 60,000 & 0.604247(65) &0.2000 &0.822(1) & 6.084(78) &$ 0.00041$(83)\\
\hline
1.85 & 0.1554600 & 105,000 & 0.594149(43) &0.2000 &0.822(1) & 7.036(87) &$-0.00013$(67)\\
\hline
1.80 & 0.1568700 & 74,500 & 0.583519(62) &0.2000 &0.824(2) & 8.42(18) &$-0.00019$(94)\\
\hline
1.75 & 0.1584100 & 60,000 & 0.572455(94) &0.2000 &0.824(2) &10.49(28) &$-0.0007$(11)\\
\hline
1.70 & 0.1599900 & 60,000 & 0.56110(11) &0.2000 &0.825(1) &15.61(56) &$ 0.0003$(12)\\
\hline
  \end{tabular}
\caption{Simulation parameters and results at $L/a$=6.}
\label{tab:simpara_L6_imp0}
\end{table}

\begin{table}
 \begin{tabular}{rccrcccr}
$\beta$ & $\kappa$ & Trajs. & plq. & $\delta\tau$ & Acc. & $g_{\rm SF}^2$ & $M$ \\
\hline
12.000 & 0.1283860 & 19,500 & 0.938309(12) &0.1000 &0.845(1) & 0.36679(56) &$-0.00014$(5)\\
\hline
8.000 & 0.1298820 & 21,500 & 0.907231(14) &0.0900 &0.909(2) & 0.5763(10) &$-0.00019$(7)\\
\hline
6.000 & 0.1314830 & 12,700 & 0.875942(18) &0.0831 &0.924(2) & 0.8137(26) &$ 0.00009$(6)\\
\hline
5.000 & 0.1328570 & 15,700 & 0.850755(16) &0.0990 &0.911(2) & 1.0250(83) &$ 0.00021$(19)\\
\hline
4.000 & 0.1351130 & 15,500 & 0.812644(20) &0.1100 &0.900(2) & 1.3886(93) &$-0.00003$(10)\\
\hline
3.000 & 0.1394120 & 29,500 & 0.748734(28) &0.1100 &0.908(2) & 2.154(11) &$-0.00002$(18)\\
\hline
2.500 & 0.1434820 & 52,100 & 0.697398(26) &0.1430 &0.843(2) & 3.000(18) &$ 0.00015$(23)\\
\hline
2.300 & 0.1458750 & 69,500 & 0.670656(21) &0.1430 &0.843(1) & 3.580(22) &$ 0.00026$(20)\\
\hline
2.200 & 0.1473293 & 69,500 & 0.655564(21) &0.1430 &0.840(1) & 4.020(31) &$ 0.00013$(23)\\
\hline
2.100 & 0.1489980 & 69,500 & 0.639127(23) &0.1430 &0.841(2) & 4.509(50) &$ 0.00028$(29)\\
\hline
2.000 & 0.1510138 & 208,900 & 0.621310(19) &0.0495 &0.834(1) & 5.452(42) &$ 0.00022$(23)\\
\hline
1.900 & 0.1533120 & 239,000 & 0.601863(21) &0.1440 &0.833(1) & 6.934(86) &$ 0.00008$(32)\\
\hline
1.825 & 0.1553190 & 268,900 & 0.586146(27) &0.1430 &0.834(1) & 9.46(15) &$-0.00021$(35)\\
\hline
1.780 & 0.1566700 & 67,300 & 0.576453(74) &0.1429 &0.834(2) &11.73(46) &$-0.00333$(90)\\
\hline
 \end{tabular}
\caption{Simulation parameters and results at $L/a$=8.}
\label{tab:simpara_L8_imp0}
\end{table}

\begin{table}
 \begin{tabular}{rccrcccr}
$\beta$ & $\kappa$ & Trajs. & plq. & $\delta\tau$ & Acc. & $g_{\rm SF}^2$ & $M$ \\
\hline
24.0000 & 0.1266100 & 57,900 & 0.969073(1) &0.0667 &0.764(2) & 0.17495(23) &$ 0.00004$(1)\\
\hline
12.0000 & 0.1280100 & 221,900 & 0.937972(1) &0.0833 &0.776(1) & 0.36828(27) &$-0.00011$(1)\\
\hline
8.0000 & 0.1295000 & 173,900 & 0.906670(2) &0.0909 &0.792(1) & 0.58292(68) &$ 0.00005$(1)\\
\hline
7.0000 & 0.1301700 & 16,100 & 0.893197(5) &0.0909 &0.804(3) & 0.6801(27) &$ 0.00011$(6)\\
\hline
6.0000 & 0.1311000 & 46,500 & 0.875162(3) &0.0909 &0.833(3) & 0.8229(28) &$ 0.00016$(4)\\
\hline
5.0000 & 0.1324800 & 60,000 & 0.849816(4) &0.1000 &0.795(2) & 1.0328(35) &$-0.00007$(4)\\
\hline
4.0000 & 0.1347120 & 62,500 & 0.811576(6) &0.0900 &0.835(4) & 1.4093(66) &$ 0.00028$(6)\\
\hline
3.5000 & 0.1364500 & 30,300 & 0.784117(10) &0.1000 &0.824(1) & 1.7137(76) &$ 0.00017$(9)\\
\hline
3.0000 & 0.1389850 & 33,500 & 0.747407(13) &0.1111 &0.779(3) & 2.205(20) &$ 0.00003$(13)\\
\hline
2.5000 & 0.1430440 & 94,500 & 0.695954(8) &0.1000 &0.842(3) & 3.128(31) &$-0.00009$(12)\\
\hline
2.3000 & 0.1454255 & 72,100 & 0.669255(10) &0.0909 &0.855(1) & 3.776(40) &$-0.00012$(13)\\
\hline
2.2000 & 0.1468380 & 109,400 & 0.654143(11) &0.0900 &0.829(5) & 4.28(49) &$ 0.00042$(17)\\
\hline
2.1000 & 0.1485305 & 121,200 & 0.637808(10) &0.0881 &0.838(5) & 4.96(82) &$ 0.00004$(17)\\
\hline
2.0000 & 0.1504620 & 60,000 & 0.619979(13) &0.1111 &0.768(2) & 5.97(19) &$-0.00015$(27)\\
\hline
1.9500 & 0.1515450 & 310,600 & 0.610434(7) &0.1111 &0.821(3) & 7.06(11) &$ 0.00065$(14)\\
\hline
1.9000 & 0.1527400 & 272,900 & 0.600522(10) &0.1111 &0.802(2) & 8.20(16) &$ 0.00029$(19)\\
\hline
1.8500 & 0.1540370 & 409,600 & 0.590204(11) &0.1111 &0.828(2) &10.92(24) &$-0.00050$(20)\\
\hline
1.8000 & 0.1554310 & 199,500 & 0.579236(23) &0.1111 &0.758(1) &17.37(96) &$ 0.00031$(31)\\
\hline
1.7500 & 0.1569430 & 442,000 & 0.567346(28) &0.1111 &0.744(4) &59.5(5.6) &$ 0.00108$(29)\\
\hline
  \end{tabular}
\caption{Simulation parameters and results at $L/a$=12.}
\label{tab:simpara_L12_imp0}
\end{table}

\begin{table}
 \begin{tabular}{rccrcccr}
$\beta$ & $\kappa$ & Trajs. & plq. & $\delta\tau$ & Acc. & $g_{\rm SF}^2$ & $M$ \\
\hline
12.0 & 0.1278810 & 17,500 & 0.937768(4) &0.0476 &0.864(2) & 0.36802(68) &$-0.00015$(2)\\
\hline
8.00 & 0.1293670 & 35,900 & 0.906361(1) &0.0432 &0.868(8) & 0.5844(18) &$ 0.00014$(2)\\
\hline
6.00 & 0.1309810 & 38,500 & 0.874761(2) &0.0432 &0.891(7) & 0.8269(34) &$-0.00010$(2)\\
\hline
5.00 & 0.1323520 & 50,300 & 0.849347(2) &0.0129 &0.871(6) & 1.0566(52) &$-0.00003$(3)\\
\hline
4.00 & 0.1345736 & 51,950 & 0.811008(3) &0.0556 &0.877(2) & 1.411(11) &$ 0.00033$(4)\\
\hline
3.00 & 0.1388563 & 32,300 & 0.746734(8) &0.0171 &0.883(4) & 2.217(33) &$ 0.00010$(6)\\
\hline
2.50 & 0.1428970 & 60,950 & 0.695290(6) &0.0175 &0.906(2) & 3.240(41) &$ 0.00023$(10)\\
\hline
2.30 & 0.1452748 & 64,700 & 0.668662(4) &0.0623 &0.873(1) & 3.924(71) &$ 0.00027$(10)\\
\hline
2.20 & 0.1467180 & 159,000 & 0.653642(4) &0.0586 &0.883(2) & 4.493(66) &$ 0.00045$(7)\\
\hline
2.10 & 0.1483940 & 150,700 & 0.637384(5) &0.0664 &0.854(1) & 5.128(86) &$-0.00004$(9)\\
\hline
2.00 & 0.1503060 & 204,200 & 0.619621(4) &0.0179 &0.852(2) & 6.56(16) &$ 0.00051$(11)\\
\hline
1.90 & 0.1525678 & 210,100 & 0.600368(6) &0.0229 &0.826(1) & 9.66(30) &$ 0.00010$(16)\\
\hline
1.87 & 0.1533100 & 160,300 & 0.594175(10) &0.0708 &0.825(1) &11.39(56) &$ 0.00029$(22)\\
\hline
  \end{tabular}
\caption{Simulation parameters and results at $L/a$=16.}
\label{tab:simpara_L16_imp0}
\end{table}

\begin{table}
 \begin{tabular}{rccrcccr}
$\beta$ & $\kappa$ & Trajs. & plq. & $\delta\tau$ & Acc. & $g_{\rm SF}^2$ & $M$ \\
\hline
8.00 & 0.1293410 & 32,500 & 0.906259(2) &0.0526 &0.852(2) & 0.5842(31) &$-0.00015$(2)\\
\hline
6.00 & 0.1309400 & 20,450 & 0.874629(2) &0.0714 &0.752(4) & 0.8300(81) &$ 0.00014$(4)\\
\hline
5.00 & 0.1323100 & 20,000 & 0.849187(3) &0.0714 &0.765(4) & 1.046(11) &$ 0.00015$(3)\\
\hline
4.00 & 0.1345380 & 20,000 & 0.810827(4) &0.0714 &0.790(4) & 1.429(23) &$ 0.00031$(5)\\
\hline
3.00 & 0.1388350 & 60,150 & 0.746527(4) &0.0769 &0.803(5) & 2.301(25) &$-0.00004$(6)\\
\hline
2.50 & 0.1429000 & 32,050 & 0.695148(6) &0.0714 &0.797(2) & 3.302(85) &$-0.00082$(13)\\
\hline
2.30 & 0.1452385 & 55,700 & 0.668468(4) &0.0625 &0.842(2) & 4.10(12) &$ 0.00049$(10)\\
\hline
2.20 & 0.1466920 & 56,000 & 0.653523(5) &0.0187 &0.860(2) & 4.503(87) &$ 0.00018$(11)\\
\hline
2.10 & 0.1483490 & 74,500 & 0.637235(5) &0.0582 &0.857(1) & 5.44(14) &$ 0.00025$(12)\\
\hline
2.00 & 0.1502770 & 142,700 & 0.619592(5) &0.1000 &0.834(4) & 6.41(13) &$ 0.00002$(11)\\
\hline
1.95 & 0.1513521 & 234,200 & 0.610169(4) &0.0625 &0.830(1) & 8.15(27) &$ 0.00035$(12)\\
\hline
1.90 & 0.1525395 & 283,950 & 0.600432(5) &0.0580 &0.854(1) & 9.50(28) &$-0.00054$(11)\\
\hline
  \end{tabular}
\caption{Simulation parameters and results at $L/a$=18.}
\label{tab:simpara_L18_imp0}
\end{table}

\begin{table}
 \begin{tabular}{rccrcccr}
$\beta$ & $\kappa$ & Trajs. & plq. & $\delta\tau$ & Acc. & $g_{\rm SF}^2$ & $M$ \\
\hline
12.0 & 0.1277800 & 13,120 & 0.937558(1) &0.0321 &0.855(7) & 0.3721(18) &$ 0.00014$(1)\\
\hline
8.00 & 0.1292790 & 28,820 & 0.906050(1) &0.0370 &0.873(3) & 0.5992(34) &$ 0.00012$(5)\\
\hline
6.00 & 0.1308900 & 33,740 & 0.874360(1) &0.0105 &0.876(9) & 0.8323(76) &$-0.00010$(2)\\
\hline
5.00 & 0.1322605 & 51,880 & 0.848876(1) &0.0369 &0.863(4) & 1.0505(95) &$ 0.00003$(2)\\
\hline
4.00 & 0.1344920 & 49,020 & 0.810465(2) &0.0455 &0.853(3) & 1.467(19) &$ 0.00003$(6)\\
\hline
3.00 & 0.1387840 & 51,360 & 0.746133(2) &0.0124 &0.877(2) & 2.337(52) &$-0.00018$(5)\\
\hline
2.30 & 0.1451880 & 52,000 & 0.668165(3) &0.0453 &0.850(2) & 4.18(13) &$ 0.00045$(6)\\
\hline
2.20 & 0.1466400 & 45,900 & 0.653255(3) &0.0147 &0.845(2) & 4.56(16) &$ 0.00008$(9)\\
\hline
2.10 & 0.1483010 & 70,350 & 0.637084(3) &0.0453 &0.845(1) & 5.50(17) &$-0.00009$(7)\\
\hline
2.00 & 0.1502180 & 214,710 & 0.619522(2) &0.0667 &0.841(1) & 6.99(23) &$ 0.00014$(6)\\
\hline
1.95 & 0.1513130 & 200,290 & 0.610297(2) &0.0450 &0.841(1) & 8.82(40) &$-0.00072$(8)\\
\hline
1.90 & 0.1524597 & 225,460 & 0.600510(3) &0.0476 &0.819(1) &10.95(57) &$ 0.00007$(9)\\
\hline
  \end{tabular}
\caption{Simulation parameters and results at $L/a$=24.}
\label{tab:simpara_L24_imp0}
\end{table}

\end{document}